\documentclass[12pt]{article}
\usepackage{enumerate}
\usepackage{amsfonts}
\usepackage{amsmath}
\usepackage{amssymb}
\usepackage{amsthm}
\usepackage{latexsym}
\usepackage{color}

\def\D{\mathcal{D}}
\def\H{\mathcal{H}}

\def\P{\mathcal{P}}
\def\SC{\mathcal{S}}
\def\S{\mathfrak{S}}
\def\PP{\mathfrak{P}}
\def\F{\mathfrak{F}}
\def\C{\mathfrak{C}}
\def\T{\mathfrak{T}}
\def\B{\mathfrak{B}}

\def\N{\mathbb{N}}

\newcommand{\supp}{\mathrm{supp}}
\newcommand{\rank}{\mathrm{rank}}

\newcommand{\id}{\mathrm{Id}}
\newcommand{\Tr}{\mathrm{Tr}}

\newcommand{\shs}{\hspace{1pt}}
\newcounter{defin}  \newcounter{lemma}  \newcounter{theorem}
\newcounter{proposition} \newcounter{corol}  \newcounter{remark} \newcounter{example}
\newenvironment{lemma}{\par\refstepcounter{lemma}     \textbf{Lemma \thelemma.} }{\rm\par}
\newenvironment{theorem}{\par\refstepcounter{theorem}     \textbf{Theorem \thetheorem.}\ }{\rm\par}
\newenvironment{proposition}{\par\refstepcounter{proposition}     \textbf{Proposition \theproposition.}\ }{\rm\par}
\newenvironment{corollary}{\par\refstepcounter{corol}     \textbf{Corollary \thecorol.} }{\rm\par}

\newenvironment{remark}{\par\refstepcounter{remark}     \textbf{Remark \theremark.}}{\rm\par}
\newenvironment{example}{\par\refstepcounter{example}     \textbf{Example \theexample.}}{\rm\par}

\begin{document}

\title{Close-to-optimal continuity bound for the von Neumann entropy and other quasi-classical applications of the Alicki-Fannes-Winter technique}

\author{M.E.~Shirokov\footnote{email:msh@mi.ras.ru}\\
Steklov Mathematical Institute, Moscow, Russia}
\date{}
\maketitle
\begin{abstract}
We consider a quasi-classical version of the Alicki-Fannes-Winter technique widely used for quantitative continuity analysis
of characteristics of quantum systems and channels. This version allows us to obtain continuity bounds under constraints
of different types for quantum states belonging to subsets of a special form that can be called "quasi-classical".

Several applications of the proposed method are described. Among others, we obtain the universal continuity bound for the von Neumann
entropy under the energy-type constraint which in the case of one-mode quantum oscillator is close to the specialized optimal continuity
bound presented recently by Becker, Datta and  Jabbour.

We obtain semi-continuity bounds for the quantum conditional entropy of quantum-classical states and for the entanglement of formation in bipartite quantum systems
with the rank/energy constraint  imposed only on one state. Semi-continuity bounds for entropic characteristics of classical random variables and classical states of a multi-mode quantum oscillator are also obtained.
\end{abstract}

\tableofcontents

\section{Introduction}

Quantitative continuity analysis of characteristics of quantum systems and channels is an important technical task which is necessary
for conducting research in various directions of quantum information theory \cite{H-SCI,Wilde}. It suffices to mention the role of the Fannes continuity bound for the quantum entropy
and the Alicki-Fannes continuity bound for the quantum conditional entropy in study of different questions that  arose when exploring the information abilities
of quantum systems and channels.

One of the universal methods of quantitative continuity analysis is the Alicki-Fannes-Winter technique. The first version of this technique was used by Alicki and Fannes to obtain a continuity bound for the quantum conditional entropy in finite-dimensional quantum systems \cite{A&F}. Then this technique was analysed and improved by different authors \cite{SR&H,M&H}, its optimal form was proposed by Winter, who applied it to get tight continuity bounds for the quantum conditional entropy and for the bipartite relative entropy of entanglement \cite{W-CB}. In a full generality this technique is described in Section 3 in \cite{QC},  its new development is proposed in the recent article  \cite{Capel}.

In this article we consider a quasi-classical version of the Alicki-Fannes-Winter technique. It  allows us to obtain continuity bounds under constraints
of different types for quantum states belonging to subsets of a special form that can be called "quasi-classical".

Two basic examples of "quasi-classical" sets of quantum states used in this article are the following:
\begin{itemize}
  \item the set of all quantum states diagonizable in some orthonormal basis;
  \item the set of classical states of a multi-mode quantum oscillator.
\end{itemize}

We describe several applications of the proposed method. In particular, it allows us to obtain universal continuity bound  and semi-continuity bound for the von Neumann
entropy under the energy-type constraint which improve the continuity bounds proposed by Winter in \cite{W-CB}.
In the case of one-mode quantum oscillator the obtained continuity bound for the von Neumann
entropy  is close to the specialized optimal continuity
bound presented  by Becker, Datta and  Jabbour in \cite{BDJ}.

We also obtain continuity bounds and semi-continuity bounds for the quantum conditional entropy
of quantum-classical states under different constraints, which allow us to derive (via the standard Nielsen-Winter
technique) continuity bounds and semi-continuity bounds for the entanglement of formation in bipartite quantum systems.
In particular, we obtain a continuity bound for the entanglement of formation under the energy-type constraint
which is essentially sharper than the previously proposed continuity bounds.

The proposed method allows us to significantly refine  continuity bounds for characteristics of composite infinite-dimensional quantum systems obtained in \cite{CBM,QC} by restricting attention to commuting states of these systems. It also allows us to refine continuity bounds for characteristics of a multi-mode quantum oscillator by restricting attention to its classical states.

Other applications considered in the article are  continuity bounds and semi-continuity bounds for entropic characteristics
of discrete random variables under constrains of different forms.

It is essential that the proposed method applied to nonnegative functions gives semi-continuity bounds for quasi-classical states
with  rank/energy constraint  imposed only on one state.

\section{Preliminaries}\label{sec2}

Let $\mathcal{H}$ be a separable Hilbert space,
$\mathfrak{B}(\mathcal{H})$ the algebra of all bounded operators on $\mathcal{H}$ with the operator norm $\|\cdot\|$ and $\mathfrak{T}( \mathcal{H})$ the
Banach space of all trace-class
operators on $\mathcal{H}$  with the trace norm $\|\!\cdot\!\|_1$. Let
$\mathfrak{S}(\mathcal{H})$ be  the set of quantum states (positive operators
in $\mathfrak{T}(\mathcal{H})$ with unit trace) \cite{H-SCI,N&Ch,Wilde}.

Denote by $I_{\mathcal{H}}$ the unit operator on a Hilbert space
$\mathcal{H}$ and by $\id_{\mathcal{\H}}$ the identity
transformation of the Banach space $\mathfrak{T}(\mathcal{H})$.

We will use the Mirsky inequality
\begin{equation}
  \sum_{i=1}^{+\infty}\vert\lambda^{\rho}_{i}-\lambda^{\sigma}_{i}\vert\leq \|\rho-\sigma\|_1\label{Mirsky-ineq+}%
\end{equation}
valid for any positive operators $\rho$ and $\sigma$ in $\T(\H)$, where  $\{\lambda^{\rho}_i\}_{i=1}^{+\infty}$
and $\{\lambda^{\sigma}_i\}_{i=1}^{+\infty}$ are  sequence
of eigenvalues of $\rho$ and $\sigma$ arranged in the non-increasing order (taking the multiplicity into account) \cite{Mirsky,Mirsky-rr}.

A finite or countable collection $\{\rho_{k}\}$ of quantum states
with a  probability distribution $\{p_{k}\}$ is called (discrete) \emph{ensemble} and denoted by $\{p_k,\rho_k\}$. The state $\bar{\rho}=\sum_{k} p_k\rho_k$ is called  the \emph{average state} of  $\{p_k,\rho_k\}$.

The \emph{von Neumann entropy} of a quantum state
$\rho \in \mathfrak{S}(\H)$ is  defined by the formula
$S(\rho)=\operatorname{Tr}\eta(\rho)$, where  $\eta(x)=-x\ln x$ if $x>0$
and $\eta(0)=0$. It is a concave lower semicontinuous function on the set~$\mathfrak{S}(\H)$ taking values in~$[0,+\infty]$ \cite{H-SCI,L-2,W}.
The von Neumann entropy satisfies the inequality
\begin{equation}\label{w-k-ineq}
S(p\rho+(1-p)\sigma)\leq pS(\rho)+(1-p)S(\sigma)+h_2(p)
\end{equation}
valid for any states  $\rho$ and $\sigma$ in $\S(\H)$ and $p\in(0,1)$, where $\,h_2(p)=\eta(p)+\eta(1-p)\,$ is the binary entropy \cite{O&P,N&Ch,Wilde}.


The \emph{quantum relative entropy} for two states $\rho$ and
$\sigma$ in $\mathfrak{S}(\mathcal{H})$ is defined as
\begin{equation*}
D(\rho\,\|\shs\sigma)=\sum_i\langle
\varphi_i\vert\,\rho\ln\rho-\rho\ln\sigma\,\vert\varphi_i\rangle,
\end{equation*}
where $\{\varphi_i\}$ is the orthonormal basis of
eigenvectors of the state $\rho$ and it is assumed that
$D(\rho\,\|\sigma)=+\infty$ if $\,\mathrm{supp}\rho\shs$ is not
contained in $\shs\mathrm{supp}\shs\sigma$ \cite{H-SCI,L-2,W}.\footnote{The support $\mathrm{supp}\rho$ of a state $\rho$ is the closed subspace spanned by the eigenvectors of $\rho$ corresponding to its positive eigenvalues.}

The \emph{quantum conditional entropy} (QCE) of a state $\rho$ of a finite-dimensional bipartite system $AB$ is defined as
\begin{equation}\label{ce-def}
S(A\vert B)_{\rho}=S(\rho)-S(\rho_{B}).
\end{equation}
The function $\rho\mapsto S(A\vert B)_{\rho}$ is  concave and
\begin{equation}\label{ce-ub}
\vert S(A\vert B)_{\rho}\vert\leq S(\rho_A)
\end{equation}
for any state $\rho$ in $\S(\H_{AB})$ \cite{H-SCI,Wilde}. By using concavity of the von Neumann entropy and inequality (\ref{w-k-ineq}) it is easy to show that
\begin{equation}\label{ce-LAA-2}
S(A\vert B)_{p\rho+(1-p)\sigma}\leq p S(A\vert B)_{\rho}+(1-p)S(A\vert B)_{\sigma}+h_2(p)
\end{equation}
for any states  $\rho$ and $\sigma$ in $\S(\H_{AB})$ and $p\in[0,1]$, where $\,h_2\,$ is the binary entropy.

Definition (\ref{ce-def}) remains valid for a state $\rho$ of an infinite-dimensional bipartite system $AB$
with finite marginal entropies
$S(\rho_A)$ and $S(\rho_B)$  (since the finiteness of $S(\rho_A)$ and $S(\rho_B)$ are equivalent to the finiteness of $S(\rho)$ and $S(\rho_B)$).
For a state $\rho$ with finite $S(\rho_A)$ and arbitrary $S(\rho_B)$ one can define the QCE
by the formula
\begin{equation}\label{ce-ext}
S(A\vert B)_{\rho}=S(\rho_{A})-D(\rho\shs\|\shs\rho_{A}\otimes\rho_{B})
\end{equation}
proposed and analysed by Kuznetsova in \cite{Kuz} (the finiteness of $S(\rho_{A})$ implies the finiteness of $D(\rho\shs\|\shs\rho_{A}\otimes\rho_{B})$). The QCE  extented by the above formula to the convex set $\,\{\rho\in\S(\H_{AB})\,\vert\,S(\rho_A)<+\infty\}\,$ possesses all basic properties of the QCE valid in finite dimensions \cite{Kuz}. In particular, it is concave and satisfies inequalities (\ref{ce-ub}) and (\ref{ce-LAA-2}).

The \emph{quantum mutual information} of a state $\,\rho\,$ in $\S(\H_{AB})$ is defined as
\begin{equation}\label{mi-d}
I(A\!:\!B)_{\rho}=D(\rho\shs\|\shs\rho_{A}\otimes
\rho_{\shs B})=S(\rho_{A})+S(\rho_{\shs B})-S(\rho),
\end{equation}
where the second formula is valid if $\,S(\rho)\,$ is finite \cite{L-mi}.
Basic properties of the relative entropy show that $\,\rho\mapsto
I(A\!:\!B)_{\rho}\,$ is a lower semicontinuous function on the set
$\S(\H_{AB})$ taking values in $[0,+\infty]$. It is well known that
\begin{equation}\label{MI-UB}
I(A\!:\!B)_{\rho}\leq 2\min\{S(\rho_A),S(\rho_B)\}
\end{equation}
for any state $\rho\in\S(\H_{AB})$ and that factor 2 in (\ref{MI-UB}) can be omitted if $\rho$ is a separable state \cite{L-mi,Wilde}.

The quantum mutual information is not convex or concave, but it satisfies the inequalities
\begin{equation}\label{MI-LAA-1}
I(A\!:\!B)_{p\rho+(1-p)\sigma} \geq p I(A\!:\!B)_{\rho}+(1-p)I(A\!:\!B)_{\sigma}-h_2(p)
\end{equation}
and
\begin{equation}\label{MI-LAA-2}
p I(A\!:\!B)_{\rho}+(1-p)I(A\!:\!B)_{\sigma}\geq I(A\!:\!B)_{p\rho+(1-p)\sigma}-h_2(p)
\end{equation}
valid for any states $\rho$ and $\sigma$ in $\S(\H_{AB})$ and  any $p\in(0,1)$  with possible values $+\infty$ in both sides \cite{CBM,QC}.

Let $H$ be a positive (semi-definite)  operator on a Hilbert space $\mathcal{H}$ (we will always assume that positive operators are self-adjoint). Denote by $\mathcal{D}(H)$ the domain of $H$. For any positive operator $\rho\in\T(\H)$ we will define the quantity $\Tr H\rho$ by the rule
\begin{equation*}
\Tr H\rho=
\left\{\begin{array}{l}
        \sup_n \Tr P_n H\rho\;\; \textrm{if}\;\;  \supp\rho\subseteq {\rm cl}(\mathcal{D}(H))\\
        +\infty\;\;\textrm{otherwise}
        \end{array}\right.
\end{equation*}
where $P_n$ is the spectral projector of $H$ corresponding to the interval $[0,n]$ and ${\rm cl}(\mathcal{D}(H))$ is the closure of $\mathcal{D}(H)$. If
$H$ is the Hamiltonian (energy observable) of a quantum system described by the space $\H$ then
$\Tr H\rho$ is the mean energy of a state $\rho$.

For any positive operator $H$ the set
$$
\C_{H,E}=\left\{\rho\in\S(\H)\,\vert\,\Tr H\rho\leq E\right\}
$$
is convex and closed (since the function $\rho\mapsto\Tr H\rho$ is affine and lower semicontinuous). It is nonempty if $E> E_0$, where $E_0$ is the infimum of the spectrum of $H$.

The von Neumann entropy is continuous on the set $\C_{H,E}$ for any $E> E_0$ if and only if the operator $H$ satisfies  the \emph{Gibbs condition}
\begin{equation}\label{H-cond}
  \Tr\, e^{-\beta H}<+\infty\quad\textrm{for all}\;\,\beta>0
\end{equation}
and the supremum of the entropy on this set is attained at the \emph{Gibbs state}
\begin{equation}\label{Gibbs}
\gamma_H(E)\doteq e^{-\beta(E) H}/\Tr e^{-\beta(E) H},
\end{equation}
where the parameter $\beta(E)$ is determined by the equation $\Tr H e^{-\beta H}=E\Tr e^{-\beta H}$ \cite{W}. Condition (\ref{H-cond}) can be valid only if $H$ is an unbounded operator having  discrete spectrum of finite multiplicity. It means, in Dirac's notation, that
\begin{equation*}
H=\sum_{k=0}^{+\infty} E_k \vert\tau_k\rangle\langle\tau_k\vert,
\end{equation*}
where
$\mathcal{T}\doteq\left\{\tau_k\right\}_{k=0}^{+\infty}$ is the orthonormal
system of eigenvectors of $H$ corresponding to the \emph{nondecreasing} unbounded sequence $\left\{E_k\right\}_{k=0}^{+\infty}$ of its eigenvalues
and \emph{it is assumed that the domain $\D(H)$ of $H$ lies within the closure $\H_\mathcal{T}$ of the linear span of $\mathcal{T}$}. In this case
\begin{equation*}
\Tr H \rho=\sum_i \lambda_i\|\sqrt{H}\varphi_i\|^2
\end{equation*}
for any positive operator $\rho$ in $\T(\H)$ with the spectral decomposition $\rho=\sum_i \lambda_i\vert\varphi_i\rangle\langle\varphi_i\vert\;$ provided that
all the vectors $\,\varphi_i\,$ lie in  $\;\D(\sqrt{H})=\{ \varphi\in\H_\mathcal{T}\,\vert \sum_{k=0}^{+\infty} E_k \vert\langle\tau_k\vert\varphi\rangle\vert^2<+\infty\}$. If at least one eigenvector of $\rho$ corresponding to a nonzero eigenvalue does not belong to the set $\D(\sqrt{H})$
then $\Tr H \rho=+\infty$.


We will use the function
\begin{equation}\label{F-def}
F_{H}(E)\doteq\sup_{\rho\in\C_{H,E}}S(\rho)=S(\gamma_H(E)).
\end{equation}
This is a strictly increasing concave function on $[E_0,+\infty)$ \cite{EC,W-CB}. It is easy to see that $F_{H}(E_0)=\ln m(E_0)$, where $m(E_0)$ is the multiplicity of $E_0$. By Proposition 1 in \cite{EC} the Gibbs condition (\ref{H-cond}) is equivalent to the following asymptotic property
\begin{equation}\label{H-cond-a}
  F_{H}(E)=o\shs(E)\quad\textrm{as}\quad E\rightarrow+\infty.
\end{equation}

For example, if $\,H=\hat{N}\doteq a^\dagger a\,$ is the number operator of a quantum oscillator then $F_H(E)=g(E)$, where
\begin{equation}\label{g-def}
  g(x)=(x+1)h_2\!\left(\frac{x}{x+1}\right)=(x+1)\ln(x+1)-x\ln x,\;\, x>0,\quad g(0)=0.
\end{equation}

We will often assume that
\begin{equation}\label{star}
  E_0\doteq\inf\limits_{\|\varphi\|=1}\langle\varphi\vert H\vert\varphi\rangle=0.
\end{equation}
In this case the concavity and nonnegativity of $F_H$ imply that (cf.\cite[Corollary 12]{W-CB})
\begin{equation}\label{W-L}
  xF_H(E/x)\leq yF_H(E/y)\quad  \forall y>x>0.
\end{equation}

\section{The Alicki-Fannes-Winter method in the quasi-classical settings}\label{sec3}

\subsection{Basic lemma}

In this subsection we describe one  general result concerning properties of a real-valued function $f$ on a convex subset $\S_0$ of $\S(\H)$ satisfying the inequalities
\begin{equation}\label{LAA-1}
  f(p\rho+(1-p)\sigma)\geq pf(\rho)+(1-p)f(\sigma)-a_f(p)
\end{equation}
and
\begin{equation}\label{LAA-2}
  f(p\rho+(1-p)\sigma)\leq pf(\rho)+(1-p)f(\sigma)+b_f(p),
\end{equation}
for all states $\rho$ and $\sigma$ in $\S_0$ and any $p\in[0,1]$, where $a_f(p)$ and $b_f(p)$ are continuous  functions on $[0,1]$  vanishing as $p\rightarrow+0$.
These  inequalities can be treated, respectively, as weakened forms of concavity and convexity. Following \cite{QC} we will call functions
satisfying both inequalities (\ref{LAA-1}) and (\ref{LAA-2}) \emph{locally almost affine} (breifly, \emph{LAA functions}), since for any such function $f$ the quantity
$\,\vert f(p\rho+(1-p)\sigma)-p f(\rho)-(1-p)f(\sigma)\vert \,$ tends to zero as $\,p\rightarrow 0^+$ uniformly on $\,\S_0\times\S_0$.


Below we consider a modification of the Alicki-Fannes-Winter method widely used for quantitative continuity analysis
of characteristics of quantum systems and channels (the brief history of appearance of this method and its most general
description can be found in  \cite[Section 3]{QC}).\smallskip

Let $\{X,\F\}$ be a measurable space and $\tilde{\omega}(x)$ a $\F$-measurable $\S(\H)$-valued function on $X$. Denote by $\P(X)$ the set of all probability measures on $X$ (more precisely, on $\{X,\F\}$). We will assume that the function $\tilde{\omega}(x)$ is integrable (in the Pettis sense \cite{P-int}) w.r.t. any measure in $\P(X)$. Consider  the set of states
\begin{equation}\label{q-set}
\mathfrak{Q}_{X,\F,\tilde{\omega}}\doteq\left\{\rho\in\S(\H)\,\left\vert\,\exists\mu_{\rho}\in\P(X):\rho=\int_X\tilde{\omega}(x)\mu_{\rho}(dx)\;\right.\right\}.
\end{equation}
We will call any measure $\mu_{\rho}$ in $\P(X)$ such that $\rho=\int_X\tilde{\omega}(x)\mu_{\rho}(dx)$ a \emph{representing measure} for a state $\rho$ in $\mathfrak{Q}_{X,\F,\tilde{\omega}}$.

We will use the total variation distance between probability measures $\mu$ and $\nu$ in $\P(X)$ defined as
\begin{equation}\label{TVD-def}
\mathrm{TV}(\mu,\nu)=\sup_{A\in\F}\vert\mu(A)-\nu(A)\vert.
\end{equation}

Consider two  examples used below:
\begin{itemize}
  \item if $X=\N$, $\F$ is the $\sigma$-algebra of all subsets of $\N$ and $\,\tilde{\omega}(n)=\vert n\rangle\langle n\vert$, where $\{\vert n\rangle\}_{n\in\N}$ is an orthonormal basic in $\H$ then $\mathfrak{Q}_{X,\F,\tilde{\omega}}$ is the set of all states in $\S(\H)$ diagonizable in the basic $\{\vert n\rangle\}$. In this case a representing measure $\mu_{\rho}$ is determined by the spectrum of $\rho$.
  \item if $X=\mathbb{C}$, $\F$ is the Borel $\sigma$-algebra on $\mathbb{C}$ and $\tilde{\omega}(z)=\vert z\rangle\langle z\vert$ -- the coherent state of a quantum oscillator
corresponding to a complex number $z$ then $\mathfrak{Q}_{X,\F,\tilde{\omega}}$ is the set of all classical states of a quantum oscillator. In this case $\mu_{\rho}(dz)=P_{\rho}(z)dz$, where $P_{\rho}$ is the $P$-function of $\rho$ \cite{Gla,Sud,IQO}.
\end{itemize}

The following lemma gives semi-continuity bound for  LAA functions on a set of quantum states having form (\ref{q-set}). It is proved by obvious modification of the Alicki-Fannes-Winter technique.

\begin{lemma}\label{g-ob} \emph{Let $\mathfrak{Q}_{X,\F,\tilde{\omega}}$ be the set defined in (\ref{q-set}) and $\S_0$ a convex subset of $\S(\H)$ with the property
}\begin{equation}\label{S-prop}
  \rho\in\S_0\cap\mathfrak{Q}_{X,\F,\tilde{\omega}}\quad \Rightarrow \quad\{\sigma\in\mathfrak{Q}_{X,\F,\tilde{\omega}}\,\vert\,\exists \varepsilon>0:\varepsilon\sigma\leq \rho\}\subseteq\S_0.
\end{equation}

\emph{Let $f$ be a function  on the set $\,\S_0$ taking values in $(-\infty,+\infty]$ that satisfies inequalities (\ref{LAA-1}) and (\ref{LAA-2}) with possible
value $+\infty$ in both sides. Let $\rho$ and $\sigma$ be states in $\,\mathfrak{Q}_{X,\F,\shs\tilde{\omega}}\cap\S_0\,$ with representing measures $\mu_{\rho}$ and $\mu_{\sigma}$ correspondingly. If $\,f(\rho)<+\infty\,$ then
\begin{equation}\label{AFW-1+}
f(\rho)-f(\sigma)\leq \varepsilon C_f(\rho,\sigma\shs\vert\shs\varepsilon)+D_f(\varepsilon),
\end{equation}
where $\varepsilon=\mathrm{TV}(\mu_{\rho},\mu_{\sigma})$, $D_f(\varepsilon)=\displaystyle(1+\varepsilon)(a_f+b_f)\!\left(\frac{\varepsilon}{1+\varepsilon}\right)$,
\begin{equation}\label{C-f}
C_f(\rho,\sigma\shs\vert\shs\varepsilon)\doteq\sup\left\{f(\varrho)-f(\varsigma)\left\vert\, \varrho,\varsigma\in\mathfrak{Q}_{X,\F,\shs\tilde{\omega}},\; \varepsilon\varrho\leq\rho,\; \varepsilon\varsigma\leq\sigma\right.\right\}
\end{equation}
and the left hand side of (\ref{AFW-1+}) may be equal to $-\infty$.}

\emph{If the function $f$ is nonnegative then inequality (\ref{AFW-1+}) holds with $C_f(\rho,\sigma\shs\vert\shs\varepsilon)$ replaced by}
\begin{equation}\label{C-f+}
C^+_f(\rho\shs\vert\shs\varepsilon)\doteq\sup\left\{f(\varrho)\,\left\vert\, \varrho\in\mathfrak{Q}_{X,\F,\tilde{\omega}},\; \varepsilon\varrho\leq\rho\right.\right\}.
\end{equation}
\end{lemma}

\begin{proof} We may assume that $f(\sigma)<+\infty$, since otherwise (\ref{AFW-1+}) holds trivially. By the condition we have
$$
2\mathrm{TV}(\mu_{\rho},\mu_{\sigma})=[\mu_{\rho}-\mu_{\sigma}]_+(X)+[\mu_{\rho}-\mu_{\sigma}]_-(X)=2\varepsilon,
$$
where $[\mu_{\rho}-\mu_{\sigma}]_+$ and $[\mu_{\rho}-\mu_{\sigma}]_-$ are the positive and negative parts
of the measure $\mu_{\rho}-\mu_{\sigma}$ (in the sense of Jordan decomposition theorem \cite{Bil}). Since $\,\mu_{\rho}(X)=\mu_{\sigma}(X)=1$, it follows from the above equality that
$\,[\mu_{\rho}-\mu_{\sigma}]_+(X)=[\mu_{\rho}-\mu_{\sigma}]_-(X)=\varepsilon$. Hence, $\nu_\pm\doteq \varepsilon^{-1}[\mu_{\rho}-\mu_{\sigma}]_\pm\in\P(X)$.
Moreover, it is easy to show, by using the definition of $\,[\mu_{\rho}-\mu_{\sigma}]_\pm$ via the Hahn decomposition of $X$, that
\begin{equation}\label{m-r}
\varepsilon\nu_+=[\mu_{\rho}-\mu_{\sigma}]_+\leq \mu_{\rho}\quad \textrm{and} \quad \varepsilon\nu_-=[\mu_{\rho}-\mu_{\sigma}]_-\leq \mu_{\sigma}.
\end{equation}

Consider the states
\begin{equation}\label{tau-s}
\tau_+=\int_X\tilde{\omega}(x)\nu_+(dx)\quad \textrm{and} \quad \tau_-=\int_X\tilde{\omega}(x)\nu_-(dx).
\end{equation}
Since the inequalities in (\ref{m-r}) imply that $\varepsilon\tau_+\leq\rho$ and $\varepsilon\tau_-\leq\sigma$, these states
belong to the set $\mathfrak{Q}_{X,\F,\tilde{\omega}}\cap\S_0$ due to condition (\ref{S-prop}). Then we have
\begin{equation}\label{omega-star}
\frac{1}{1+\varepsilon}\,\rho+\frac{\varepsilon}{1+\varepsilon}\,\tau_-=\omega_*=
\frac{1}{1+\varepsilon}\,\sigma+\frac{\varepsilon}{1+\varepsilon}\,\tau_+,
\end{equation}
where $\omega_*$ is a state in $\S_0$. By the finiteness of $f(\rho)$ and $f(\sigma)$ the operator inequalities $\,\varepsilon\tau_+\leq\rho\,$ and $\,\varepsilon\tau_-\leq\sigma\,$ along with inequalities (\ref{LAA-1}) and (\ref{LAA-2})
imply the finiteness of $f(\tau_+)$, $f(\tau_-)$ and $f(\omega_*)$.

By applying inequalities (\ref{LAA-1}) and (\ref{LAA-2}) to the decompositions of $\omega_*$  in (\ref{omega-star}) we  obtain
$$
(1-p)(f(\rho)-f(\sigma))\leq p
(f(\tau_+)-f(\tau_-))+a_f(p)+b_f(p),
$$
where $p=\varepsilon/(1+\varepsilon)$. If follows that
\begin{equation*}
f(\rho)-f(\sigma)\leq \varepsilon(f(\tau_+)-f(\tau_-))+D_f(\varepsilon).
\end{equation*}
Since $\varepsilon\tau_+\leq\rho$ and $\varepsilon\tau_-\leq\sigma$, this implies inequality (\ref{AFW-1+}).

The last claim of the lemma is obvious.
\end{proof}


\begin{remark}\label{g-ob-r} In general, the condition $\mathrm{TV}(\mu_{\rho},\mu_{\sigma})=\varepsilon$ in
Lemma \ref{g-ob} can not be replaced by the condition $\mathrm{TV}(\mu_{\rho},\mu_{\sigma})\leq\varepsilon$, since
the functions $\varepsilon \mapsto C_f(\rho,\sigma\shs\vert\shs\varepsilon)$ and $\varepsilon \mapsto C^+_f(\rho\shs\vert\shs\varepsilon)$ may be decreasing and, hence, special arguments are required to show that the r.h.s. of (\ref{AFW-1+}) is a nondecreasing function of $\varepsilon$.
\end{remark}\smallskip

\begin{remark}\label{S-prop-r} Condition (\ref{S-prop}) in Lemma \ref{g-ob} can be replaced by the condition
\begin{equation}\label{S-prop+}
  \rho,\sigma\in\S_0\cap\mathfrak{Q}_{X,\F,\tilde{\omega}}\quad \Rightarrow \quad \tau_+,\tau_-\in\S_0,
\end{equation}
where $\tau_+$ and $\tau_-$ are the states defined in (\ref{tau-s}) via the representing measures $\mu_{\rho}$ and $\mu_{\sigma}$. In this case
one should correct the definitions of $C_f(\rho,\sigma\shs\vert\shs\varepsilon)$ and $C^+_f(\rho\shs\vert\shs\varepsilon)$
by replacing $\mathfrak{Q}_{X,\F,\shs\tilde{\omega}}$ in (\ref{C-f}) and (\ref{C-f+}) with $\mathfrak{Q}_{X,\F,\shs\tilde{\omega}}\cap\S_0$.
\end{remark}

\subsection{General results}

Among characteristics of a $n$-partite
finite-dimensional quantum system $A_{1}...A_{n}$ there are many that satisfy inequalities (\ref{LAA-1}) and (\ref{LAA-2})
with the functions $a_f$ and $b_f$ proportional to the binary entropy (defined after (\ref{w-k-ineq})) and the inequality
\begin{equation}\label{Cm}
-c^-_f C_m(\rho)\leq f(\rho)\leq c^+_f C_m(\rho),\quad C_m(\rho)=\sum_{k=1}^m S(\rho_{A_k}),\;\; m\leq n,
\end{equation}
for any state $\rho$ in $\S(\H_{A_{1}..A_{n}})$, where $c^-_f$ and $c^+_f$ are nonnegative numbers.

Following \cite{CBM,QC} introduce the class $L_n^m(C,D\vert\S_0)$, $m\leq n$, of functions on a convex subset $\S_0$ of $\S(\H_{A_{1}..A_{n}})$
satisfying inequalities  (\ref{LAA-1}) and (\ref{LAA-2})
with $a_f(p)=d_f^-h_2(p)$ and $b_f(p)=d_f^+h_2(p)$ and inequality (\ref{Cm}) for any states in $\S_0$ with the parameters $c^{\pm}_f$ and  $d^{\pm}_f$ such that $c^-_f+c^+_f=C$ and $d^-_f+d^+_f=D$.

If $A_1$,...,$A_n$ are arbitrary infinite-dimensional quantum systems then we define the classes $L_n^m(C,D\vert\S_0)$, $m\leq n$, by the same rule assuming that all functions in $L_n^m(C,D\vert \S_0)$ take values in $(-\infty,+\infty]$ on $\S_0$  and that all the
inequalities in (\ref{LAA-1}), (\ref{LAA-2}) and (\ref{Cm}) hold with possible infinite values in both sides.

We also introduce the class $\widehat{L}^{m}_n(C,D\vert \S_0)$ obtained by adding to the class $L^{m}_n(C,D\vert \S_0)$ all functions of the form
$$
f(\rho)=\inf_{\lambda}f_{\lambda}(\rho)\quad \textrm{and} \quad f(\rho)=\sup_{\lambda}f_{\lambda}(\rho),
$$
where $\{f_{\lambda}\}$ is some family of functions in $L^{m}_n(C,D\vert \S_0)$.

If $\S_0=\S_m(\H_{A_{1}..A_{n}})$, where
\begin{equation*}
\mathfrak{S}_m(\mathcal{H}_{A_1..A_n})\doteq\left\{\rho\in\mathfrak{S}(\mathcal{H}_{A_1..A_n})\,\vert\,S(\rho_{A_1}),..., S(\rho_{A_m})<+\infty\shs\right\},
\end{equation*}
then we will denote the classes $L^{m}_n(C,D\vert \S_0)$ and $\widehat{L}^{m}_n(C,D\vert \S_0)$
by $L^{m}_n(C,D)$ and $\widehat{L}^{m}_n(C,D)$ for brevity.\footnote{The necessity to consider the case $\,\S_0\neq\S_m(\H_{A_{1}..A_{n}})$ will be shown in Section 4.}

For example, the von Neumann entropy belongs to the class $L_1^1(1,1\vert\S(\H))$, while the (extended) quantum conditional entropy $S(A_1\vert A_2)$ (defined in (\ref{ce-ext})) lies in $L_2^1(2,1)\doteq L_2^1(2,1\vert\S_1(\H_{A_1A_2}))$. This follows from the concavity of these characteristics and inequalities (\ref{w-k-ineq}), (\ref{ce-ub}) and (\ref{ce-LAA-2}). The nonnegativity of $S(A_1\vert A_2)$
on the convex set $\S^1_{\rm sep}$ of separable states in $\S_1(\H_{A_1A_2})$ implies that this function also belongs to the class $L_2^1(1,1\vert \S^1_{\rm sep})$. More complete list of characteristics of quantum systems belonging the classes $L^{m}_n(C,D)$ and $\widehat{L}^{m}_n(C,D)$
can be found in \cite{CBM,QC}.

Now we apply Lemma \ref{g-ob} in Section 3.1 to obtain  continuity bounds for functions from the classes
$\widehat{L}_n^{m}(C,D\vert \S_0)$ under the rank constraint on the marginal states corresponding to the subsystems $A_1,...,A_m$.\smallskip

\begin{theorem}\label{main-1} \emph{Let  $\mathfrak{Q}_{X,\F,\tilde{\omega}}$ be the set of states in $\S(\H_{A_1...A_n})$ defined in (\ref{q-set}) via  some $\S(\H_{A_1...A_n})$-valued function $\tilde{\omega}(x)$ and $\S_0$ a convex subset of $\S(\H_{A_1...A_n})$ possessing property (\ref{S-prop}).}\smallskip

\emph{\noindent\emph{A)} If $f$ is a  function  in $\widehat{L}_n^{m}(C,D\vert \S_0)$ then
\begin{equation}\label{main+1}
    \vert f(\rho)-f(\sigma)\vert \leq C\varepsilon \ln d_m(\rho,\sigma)+Dg(\varepsilon)
\end{equation}
for any states $\rho$ and  $\sigma$ in $\mathfrak{Q}_{X,\F,\tilde{\omega}}\cap\S_0$ such that $\,\mathrm{TV}(\mu_{\rho},\mu_{\sigma})\leq \varepsilon$ and $\,d_m(\rho,\sigma)\doteq\max\left\{\prod_{k=1}^{m}\rank\rho_{A_k}, \prod_{k=1}^{m}\rank\sigma_{A_k}\right\}$ is finite, where $\mu_{\rho}$ and $\mu_{\sigma}$ are measures in $\P(X)$ representing the states $\rho$ and $\sigma$ and $g$ is the function defined in (\ref{g-def}).}\smallskip

\emph{\noindent\emph{B)} If $f$ is a nonnegative function in $\widehat{L}_n^{m}(C,D\vert \S_0)$ and $\rho$ is a state in $\,\mathfrak{Q}_{X,\F,\tilde{\omega}}\cap\S_0\,$ such that $\,d_m(\rho)\doteq\prod_{k=1}^{m}\rank\rho_{A_k}$ is finite then
\begin{equation}\label{main++1}
    f(\rho)-f(\sigma)\leq C\varepsilon \ln d_m(\rho)+Dg(\varepsilon)
\end{equation}
for any state $\sigma$ in $\mathfrak{Q}_{X,\F,\tilde{\omega}}\cap\S_0$ such that $\,\mathrm{TV}(\mu_{\rho},\mu_{\sigma})\leq \varepsilon$ (the l.h.s. of (\ref{main++1}) may be equal to $-\infty$).}
\end{theorem}\smallskip

\begin{remark}\label{main-1-r}
It is essential that in part B of Theorem \ref{main-1}
we impose no restrictions on the state $\sigma$ other than the requirement $\mathrm{TV}(\mu_{\rho},\mu_{\sigma})\leq \varepsilon$.
\end{remark}\smallskip

\begin{remark}\label{S-prop-r+}
Both claims of Theorem \ref{main-1} remain valid if $\S_0$ is \emph{any} convex subset of $\S(\H_{A_1...A_n})$
provided that condition (\ref{S-prop+}) holds for the states $\rho$ and $\sigma$. This follows from
the proof of this theorem and Remark \ref{S-prop-r}.
\end{remark}\smallskip

\begin{proof} In the proofs of both parts of the theorem we may assume that $f$ is a function in $L_n^{m}(C,D\vert \S_0)$ and that
$\mathrm{TV}(\mu_{\rho},\mu_{\sigma})=\varepsilon$. Indeed,  the expressions in r.h.s. of (\ref{main+1}) and  (\ref{main++1}) are  non-decreasing
functions of $\varepsilon$ and depend only on the parameters $C$ and $D$ and the characteristics of the states $\rho$ and $\sigma$.

A) Note first that the condition $d_m(\rho,\sigma)<+\infty$ implies that $f(\rho),f(\sigma)<+\infty$ by inequality (\ref{Cm}).
Since $\,\rank\varrho_{A_k}\leq\rank\rho_{A_k}$ and $\,\rank\varsigma_{A_k}\leq\rank\sigma_{A_k}$, $k=\overline{1,m}$,
for any states $\varrho$ and $\varsigma$ in $\mathfrak{Q}_{X,\F,\shs\tilde{\omega}}$ such that $\varepsilon\varrho\leq\rho$ and $\varepsilon\varsigma\leq\sigma$, inequality (\ref{Cm}) implies that
\begin{equation}\label{p-ineq-1}
\vert f(\varrho)-f(\varsigma)\vert \leq C\ln d_m(\rho,\sigma)
\end{equation}
for any such states $\varrho$ and $\varsigma$.  So, the quantities $C_f(\rho,\sigma\shs\vert \shs\varepsilon)$ and $C_f(\sigma,\rho\shs\vert \shs\varepsilon)$ defined in (\ref{C-f}) do not exceed
the r.h.s. of (\ref{p-ineq-1}). Thus, by applying Lemma \ref{g-ob} twice we obtain (\ref{main+1}).\smallskip

B) The condition $d_m(\rho)<+\infty$ implies that $f(\rho)<+\infty$ by inequality (\ref{Cm}). So, this assertion follows from the last claim of Lemma \ref{g-ob}, since $\,\rank\varrho_{A_k}\leq\rank\rho_{A_k}<+\infty$, $k=\overline{1,m}$,
for any state $\varrho$ in $\mathfrak{Q}_{X,\F,\shs\tilde{\omega}}$ such that $\varepsilon\varrho\leq\rho$. Thus,
inequality (\ref{Cm}) implies that
\begin{equation}\label{p-ineq}
f(\varrho)\leq C\ln d_m(\rho)
\end{equation}
for any such state $\varrho$. So, the quantity $C^+_f(\rho\shs\vert \shs\varepsilon)$ defined in (\ref{C-f+}) does not exceed
the r.h.s. of (\ref{p-ineq}).
\end{proof}

Now we apply Lemma \ref{g-ob} in Section 3.1 to obtain  continuity bounds for functions from the classes
$\widehat{L}_n^{m}(C,D\vert \S_0)$ under the constraint  corresponding to the positive operator $H_m\otimes I_{A_{m+1}}\otimes...\otimes I_{A_n}$, where
\begin{equation}\label{H-m}
H_{m}=H_{A_1}\otimes I_{A_2}\otimes...\otimes I_{A_m}+\cdots+I_{A_1}\otimes... \otimes I_{A_{m-1}}\otimes H_{A_m}
\end{equation}
is a positive operator on the space $\H_{A_1...A_m}$ determined by positive operators $H_{A_1}$,....,$H_{A_m}$ on the spaces $\H_{A_1}$,....,$\H_{A_m}$ (it is assumed that $H_1=H_{A_1}$). It is essential that the operator $H_{m}$ satisfies condition (\ref{H-cond}) if all the operators $H_{A_1}$,....,$H_{A_m}$  satisfy this condition (this follows from the equivalence of (\ref{H-cond}) and (\ref{H-cond-a}), see the proof of \cite[Lemma 4]{CBM}). Note that $\Tr H_{m}\rho=\sum_{k=1}^{m}\Tr H_{A_k}\rho_{A_k}$ for any $\rho$ in $\S(\H_{A_1..A_m})$.
We will use the function
\begin{equation}\label{F-H-m}
F_{H_{m}}(E)=\sup\shs\{ S(\rho)\,\vert \,\rho\in\S(\H_{A_1..A_m}),\,\Tr H_{m}\rho\leq E \}.
\end{equation}
If all the operators $H_{A_1}$,....,$H_{A_m}$  are isometrically equivalent to each other then  $F_{H_{m}}(E)=mF_{H_{\!A_1}}(E/m)$, where $F_{H_{\!A_1}}$ is the function defined in (\ref{F-def}) \cite[Lemma 4]{CBM}.\smallskip

\begin{theorem}\label{main} \emph{Let $H_{\!A_1}$,...,$H_{\!A_m}$ be positive operators on the Hilbert spaces $\H_{A_1}$,....,$\H_{A_m}$ satisfying condition (\ref{H-cond})
and (\ref{star}) and $F_{H_{m}}$ the function defined in (\ref{F-H-m}). Let  $\mathfrak{Q}_{X,\F,\tilde{\omega}}$ be the set of states in $\S(\H_{A_1...A_n})$ defined in (\ref{q-set}) via  $\S(\H_{A_1...A_n})$-valued  function $\tilde{\omega}(x)$ and $\S_0$ a convex subset of $\S(\H_{A_1...A_n})$ with property (\ref{S-prop}).
}
\smallskip

\emph{\noindent\emph{A)} If $f$ is a function in $\widehat{L}_n^{m}(C,D\vert \S_0)$ then
\begin{equation}\label{main+}
    \vert f(\rho)-f(\sigma)\vert \leq C\varepsilon F_{H_{m}}(mE/\varepsilon)+Dg(\varepsilon)
\end{equation}
for any states $\rho$ and $\sigma$ in $\,\mathfrak{Q}_{X,\F,\tilde{\omega}}\cap\S_0\,$ such that $\,\sum_{k=1}^{m}\Tr H_{A_k}\rho_{A_k},\,\sum_{k=1}^{m}\Tr H_{A_k}\sigma_{A_k}\leq mE$  and  $\,\mathrm{TV}(\mu_{\rho},\mu_{\sigma})\leq \varepsilon$, where $\mu_{\rho}$ and $\mu_{\sigma}$ are measures in $\P(X)$ representing the states $\rho$ and $\sigma$ and $g$ is the function defined in (\ref{g-def}).}\smallskip

\emph{\noindent\emph{B)} If $f$ is a nonnegative function in $\widehat{L}_n^{m}(C,D\vert \S_0)$ and $\rho$ is a state in $\,\mathfrak{Q}_{X,\F,\tilde{\omega}}\cap\S_0\,$ such that $\,\sum_{k=1}^{m}\Tr H_{A_k}\rho_{A_k}\leq mE\,$  then
\begin{equation}\label{main++}
    f(\rho)-f(\sigma)\leq C\varepsilon F_{H_{m}}(mE/\varepsilon)+Dg(\varepsilon)
\end{equation}
for any state $\sigma$ in $\mathfrak{Q}_{X,\F,\tilde{\omega}}\cap\S_0$ such that $\,\mathrm{TV}(\mu_{\rho},\mu_{\sigma})\leq \varepsilon$ (the l.h.s. of (\ref{main++}) may be equal to $-\infty$). If the set $\mathfrak{Q}_{X,\F,\tilde{\omega}}$ consists of commuting states then  (\ref{main++}) holds with $mE$ replaced by $\,mE-\sum_{k=1}^{m}\Tr H_{A_k}\rho^{\shs\varepsilon}_{A_k}$,
where $\,\rho^{\shs\varepsilon}\doteq[\rho-\varepsilon I_{A_1...A_n}]_+$ is the positive part of the Hermitian operator $\,\rho-\varepsilon I_{A_1...A_n}$.}
\end{theorem}\smallskip

\begin{remark}\label{main-r-1}
Since the operator $H_m$ satisfies conditions (\ref{H-cond}) and (\ref{star}), the equivalence of (\ref{H-cond}) and (\ref{H-cond-a}) and inequality (\ref{W-L}) show that the
r.h.s. of (\ref{main+}) and  (\ref{main++}) are non-decreasing functions of $\varepsilon$  tending to zero as $\varepsilon\to0$.
\end{remark}\smallskip

\begin{remark}\label{main-r-2}
It is essential that in part B of Theorem \ref{main} we impose no restrictions on the state $\sigma$ other than the requirement $\mathrm{TV}(\mu_{\rho},\mu_{\sigma})\leq \varepsilon$.
\end{remark}\smallskip

\begin{remark}\label{S-prop-r++}
Both claims of Theorem \ref{main} remain valid if $\S_0$ is \emph{any} convex subset of $\S(\H_{A_1...A_n})$
provided that condition (\ref{S-prop+}) holds for the states $\rho$ and $\sigma$. This follows from
the proof of this theorem and Remark \ref{S-prop-r}.
\end{remark}\smallskip

\begin{proof}  In the proofs of both parts of the theorem we may assume that $f$ is a function in $L_n^{m}(C,D\vert \S_0)$ and that
$\mathrm{TV}(\mu_{\rho},\mu_{\sigma})=\varepsilon$. Indeed, by Remark \ref{main-r-1} the expressions in r.h.s. of (\ref{main+}) and  (\ref{main++}) are non-decreasing
functions of $\varepsilon$ and depend only on the parameters $C$ and $D$ and the characteristics of the states $\rho$ and $\sigma$.\smallskip

A) Since $\Tr H_{m}[\vartheta_{A_1}\otimes...\otimes\vartheta_{A_m}]=\sum_{k=1}^m \Tr H_{A_k}\vartheta_{A_k}$, we have
$$
\sum_{k=1}^m S(\vartheta_{A_k})=S(\vartheta_{A_1}\otimes...\otimes\vartheta_{A_m})\leq F_{H_{m}}(mE)
$$
for any state $\vartheta\in \S_0$ such that $\Tr H_{m}\vartheta_{A_1..A_m}=\sum_{k=1}^m\Tr H_{A_k}\vartheta_{A_k}\leq mE$. Hence, for any such state $\vartheta$  inequality (\ref{Cm}) implies that
\begin{equation}\label{F-p-2+}
-c_f^-F_{H_{m}}(mE)\leq f(\vartheta)\leq c_f^+F_{H_{m}}(mE).
\end{equation}
It follows, in particular, that $f(\rho)$ and $f(\sigma)$ are finite.

Assume that $\varrho$ and $\varsigma$ are states in $\mathfrak{Q}_{X,\F,\shs\tilde{\omega}}$ such that $\varepsilon\varrho\leq\rho$ and $\varepsilon\varsigma\leq\sigma$. Then
\begin{equation}\label{en-est}
\varepsilon\sum_{k=1}^m \Tr H_{A_k}\varrho_{A_k}\leq\sum_{k=1}^m \Tr H_{A_k}\rho_{A_k}\leq mE
\end{equation}
and
$$
\varepsilon\sum_{k=1}^m \Tr H_{A_k}\varsigma_{A_k}\leq\sum_{k=1}^m \Tr H_{A_k}\sigma_{A_k}\leq mE.
$$
Thus, it follows from (\ref{F-p-2+}) that
$$
\vert f(\varrho)-f(\varsigma)\vert \leq CF_{H_{m}}(mE/\varepsilon).
$$
So, the quantities $C_f(\rho,\sigma\shs\vert \shs\varepsilon)$ and $C_f(\sigma,\rho\shs\vert \shs\varepsilon)$ defined in (\ref{C-f})  do not exceed
$CF_{H_{m}}(mE/\varepsilon)$. Thus, by applying Lemma \ref{g-ob} twice we obtain (\ref{main+}).\smallskip

B) The main assertion of part B follows from the last claim of Lemma \ref{g-ob}, since the arguments from the proof of part A show
that $f(\rho)<+\infty$ and that
\begin{equation}\label{p-ineq-2}
f(\varrho)\leq CF_{H_{m}}(mE/\varepsilon)
\end{equation}
for any state $\varrho$ in $\mathfrak{Q}_{X,\F,\shs\tilde{\omega}}$ such that $\varepsilon\varrho\leq\rho$.
So, the quantity $C^+_f(\rho\shs\vert \shs\varepsilon)$ defined in (\ref{C-f+}) does not exceed
the r.h.s. of (\ref{p-ineq-2}).

If a state $\varrho$ commutes with the state $\rho$ then the condition $\,\varepsilon\varrho\leq\rho$
implies that
$$
\varepsilon\varrho\leq \rho-[\rho-\varepsilon I_{A_1...A_n}]_+,
$$
since $\varrho\leq I_{A_1...A_n}$. By incorporating this  to the estimates in (\ref{en-est}) it is easy to prove the last claim of part B.
\end{proof}

Theorems \ref{main-1} and \ref{main} allows us to refine the continuity bounds for functions from the classes $\widehat{L}_n^{m}(C,D)$
presented in \cite{CBM,QC} in the case of commuting states, i.e. such states $\rho$ and $\sigma$ in $\,\S(\H_{A_1..A_n})$ that
$$
[\rho,\sigma]\doteq \rho\,\sigma-\sigma\rho=0.
$$
Indeed, since for any commuting states $\rho$ and $\sigma$ in $\,\S(\H_{A_1..A_n})$ there exists an orthonormal
basis $\{\vert k\rangle\}_{k\in\N}$ in $\H_{A_1..A_n}$ in which these states are diagonizable, they belong to the set
$\mathfrak{Q}_{X,\F,\tilde{\omega}}$, where $X=\N$, $\F$ is the $\sigma$-algebra of all subsets of $\N$ and $\tilde{\omega}(k)=\vert k\rangle\langle k\vert$.
In this case the set $\P(X)$ can be identified with the set of probability distributions on $\N$, the representing measures $\mu_{\rho}$ and $\mu_{\sigma}$
correspond to the probability distributions formed by the eigenvalues of $\rho$ and $\sigma$ (taken the multiplicity into account) and
$\mathrm{TV}(\mu_{\rho},\mu_{\sigma})=\textstyle\frac{1}{2}\|\rho-\sigma\|_1$.\smallskip

Thus, Theorem \ref{main-1} implies the following\smallskip

\begin{corollary}\label{main-1-c} \emph{Let  $\S_0$ be a convex subset of $\S(\H_{A_1...A_n})$ with the property\footnote{This is a  weakened form of the $\Delta$-invariance property \cite[Section 3]{QC}.}
\begin{equation}\label{S-prop++}
  \frac{[\rho-\sigma]_\pm}{\Tr[\rho-\sigma]_\pm}\in\S_0\;\textrm{ for any }\rho\textrm{ and }\sigma\textrm{ in }\,\S_0\,\textrm{ such that }\,[\rho,\sigma]=0\,\textrm{ and }\, \rho\neq\sigma,
\end{equation}
where $[\rho-\sigma]_\pm$ are the positive and negative parts of the Hermitian operator  $\rho-\sigma$.}\smallskip

\noindent A) \emph{If $f$ is a  function  in $\widehat{L}_n^{m}(C,D\vert \S_0)$  then
\begin{equation*}
    \vert f(\rho)-f(\sigma)\vert \leq C\varepsilon \ln d_m(\rho,\sigma)+Dg(\varepsilon)
\end{equation*}
for any states $\rho$ and  $\sigma$ in $\,\S_0$ s.t. $\,d_m(\rho,\sigma)\doteq\max\left\{\prod_{k=1}^{m}\rank\rho_{A_k}, \prod_{k=1}^{m}\rank\sigma_{A_k}\right\}$ is finite, $\,[\rho,\sigma]=0\,$ and $\,\frac{1}{2}\|\rho-\sigma\|_1\leq \varepsilon$.}\smallskip

\noindent B) \emph{If $f$ is a nonnegative function in $\widehat{L}_n^{m}(C,D\vert \S_0)$ and $\rho$ is a state in $\,\S(\H_{A_1..A_n})$ such that $\,d_m(\rho)\doteq\prod_{k=1}^{m}\rank\rho_{A_k}$ is finite then
\begin{equation}\label{main++1-c}
    f(\rho)-f(\sigma)\leq C\varepsilon \ln d_m(\rho)+Dg(\varepsilon)
\end{equation}
for any state $\sigma$ in $\,\S_0$ such that  $\,[\rho,\sigma]=0\,$ and $\frac{1}{2}\|\rho-\sigma\|_1\leq \varepsilon$ (the l.h.s. of (\ref{main++1-c}) may be equal to $-\infty$).}
\end{corollary}\smallskip

\begin{proof}  By the observation before the corollary all its  claims  follow from Theorem \ref{main-1}
and  Remark \ref{S-prop-r+}, since property (\ref{S-prop++}) guarantees the validity of
condition  (\ref{S-prop+}) for any commuting states in $\S_0$ as in this case $\,\tau_{\pm}=[\rho-\sigma]_\pm/\Tr[\rho-\sigma]_\pm$.
\end{proof}

\smallskip

Theorem \ref{main} implies the following\smallskip

\begin{corollary}\label{main-c} \emph{Let $\S_0$ be a convex subset of $\,\S(\H_{A_1...A_n})$ with property (\ref{S-prop++}). Let $H_{\!A_1}$,...,$H_{\!A_m}$ be positive operators on the Hilbert spaces $\H_{A_1}$,....,$\H_{A_m}$ satisfying conditions (\ref{H-cond})
and (\ref{star}) and $F_{H_{m}}$ the function defined in (\ref{F-H-m}).}\smallskip

\noindent A) \emph{If $f$ is a function in $\widehat{L}_n^{m}(C,D\vert \S_0)$ then
\begin{equation*}
    \vert f(\rho)-f(\sigma)\vert \leq C\varepsilon F_{H_{m}}(mE/\varepsilon)+Dg(\varepsilon)
\end{equation*}
for any  states $\rho$ and $\sigma$ in $\,\S(\H_{A_1..A_n})$ s.t. $\,\sum_{k=1}^{m}\Tr H_{A_k}\rho_{A_k},\,\sum_{k=1}^{m}\Tr H_{A_k}\sigma_{A_k}\leq mE\,$,
$\,[\rho,\sigma]=0$ and $\,\frac{1}{2}\|\rho-\sigma\|_1\leq\varepsilon$.}\smallskip

\noindent B) \emph{If $f$ is a nonnegative function in $\widehat{L}_n^{m}(C,D\vert \S_0)$ and $\rho$ is a state in $\,\S(\H_{A_1..A_n})$ such that $\,\sum_{k=1}^{m}\Tr H_{A_k}\rho_{A_k}\leq mE\,$  then
\begin{equation}\label{main++c}
    f(\rho)-f(\sigma)\leq C\varepsilon F_{H_{m}}((mE-E_{\varepsilon}(\rho))/\varepsilon)+Dg(\varepsilon)\leq C\varepsilon F_{H_{m}}(mE/\varepsilon)+Dg(\varepsilon)
\end{equation}
for any state $\sigma$ in $\,\S_0$ such that $\,[\rho,\sigma]=0$ and $\,\frac{1}{2}\|\rho-\sigma\|_1\leq \varepsilon$,
where\footnote{$\,[\rho-\varepsilon I_{A_1...A_n}]_+$ is the positive part of the Hermitian operator $\,\rho-\varepsilon I_{A_1...A_n}$.}
$$
E_{\varepsilon}(\rho)=\sum_{k=1}^{m}\Tr H_{A_k}\langle[\rho-\varepsilon I_{A_1...A_n}]_+\rangle_{A_k}
$$
and  the left hand sides of (\ref{main++c}) may be equal to $-\infty$.}
\end{corollary}

\smallskip

\begin{proof} By the observation before Corollary \ref{main-1-c} all the claims of this corollary follow from Theorem \ref{main}
and  Remark \ref{S-prop-r++}. It suffices to note that property (\ref{S-prop++}) guarantees the validity of
condition  (\ref{S-prop+}) for any commuting states $\rho$ and $\sigma$ in $\S_0$, since in this case $\,\tau_{\pm}=[\rho-\sigma]_\pm/\Tr[\rho-\sigma]_\pm$.
\end{proof}

\smallskip

\noindent\textbf{Note:} Since $E_{\varepsilon}(\rho)$ tends to $E(\rho)\doteq\sum_{k=1}^{m}\Tr H_{A_k}\rho_{A_k}$ as $\,\varepsilon\to0$, the first estimate in (\ref{main++c}) is essentially  sharper than the second one provided that $\,mE=E(\rho)$ and $\varepsilon\ll1$.

\smallskip

\begin{example}\label{c-1-e}  Let $f=I(A_1\!:\!A_2)$ be the quantum mutual
information defined in (\ref{mi-d}). This is a function on the whole space $\S(\H_{A_1A_2})$  taking values in $[0,+\infty]$ and satisfying the inequalities (\ref{MI-LAA-1}) and (\ref{MI-LAA-2}) with possible value $+\infty$ in both sides. This and upper bound (\ref{MI-UB}) show that $I(A_1\!:\!A_2)$ belongs to the class
$L_2^1(2,2\vert \S(\H_{A_1A_2}))$.

Thus, Corollary \ref{main-1-c}B implies that
\begin{equation}\label{I-1-1}
I(A_1\!:\!A_2)_\rho-I(A_1\!:\!A_2)_\sigma\leq 2\varepsilon \ln \rank\rho_{A_1}+2g(\varepsilon)
\end{equation}
for any commuting states $\rho$ and $\sigma$ in $\,\S(\H_{A_1A_2})$ such that $\frac{1}{2}\|\rho-\sigma\|_1\leq \varepsilon$.

The semi-continuity bound (\ref{I-1-1}) and the symmetry arguments show that
\begin{equation}\label{I-2-1}
-(2\varepsilon \ln \rank\sigma_{A_x}+2g(\varepsilon))\leq I(A_1\!:\!A_2)_\rho-I(A_1\!:\!A_2)_\sigma\leq 2\varepsilon \ln \rank\rho_{A_1}+2g(\varepsilon)
\end{equation}
for any commuting states $\rho$ and $\sigma$ in $\,\S(\H_{A_1A_2})$ such that $\frac{1}{2}\|\rho-\sigma\|_1\leq \varepsilon$, where $x$ is either $1$ or $2$.
The continuity bound (\ref{I-2-1}) gives \emph{more accurate} estimates than the universal continuity bound for $I(A_1\!:\!A_2)$ presented in \cite[Proposition 6]{QC}. Indeed, by using the last continuity bound one can only show that
$$
\vert I(A_1\!:\!A_2)_\rho-I(A_1\!:\!A_2)_\sigma\vert \leq 2\varepsilon \ln (\rank\rho_{A_1}+\rank\sigma_{A_1})+2g(\varepsilon)
$$
for any states $\rho$ and $\sigma$  such that $\frac{1}{2}\|\rho-\sigma\|_1\leq \varepsilon$.

The possibility to use in (\ref{I-2-1}) the ranks of marginal states corresponding to different subsystems (the case $x=2$) gives additional
flexibility in getting  more accurate estimate for the quantity $I(A_1\!:\!A_2)_\rho-I(A_1\!:\!A_2)_\sigma$.

Let $H$ be any positive operator on $\H_{A_1}$ satisfying conditions (\ref{H-cond})
and (\ref{star}). Corollary \ref{main-c}B implies that
\begin{equation}\label{I-1-2}
I(A_1\!:\!A_2)_\rho-I(A_1\!:\!A_2)_\sigma\leq 2\varepsilon F_{H}(E/\varepsilon)+2g(\varepsilon)
\end{equation}
for any state $\rho$ in $\,\S(\H_{A_1A_2})$ such that $\,\Tr H\rho_{A_1}\leq E\,$ and arbitrary state  $\sigma$ in $\,\S(\H_{A_1A_2})$ such that
$[\rho,\sigma]=0$ and $\frac{1}{2}\|\rho-\sigma\|_1\leq \varepsilon$. The semi-continuity bound (\ref{I-1-2}) can be refined by replacing $E$ with $E-\Tr H\langle[\rho-\varepsilon I_{A_1A_2}]_+\rangle_{A_1}$.

It follows from (\ref{I-1-2}) that
\begin{equation}\label{I-2-2}
\vert I(A_1\!:\!A_2)_\rho-I(A_1\!:\!A_2)_\sigma\vert \leq 2\varepsilon F_{H}(E/\varepsilon)+2g(\varepsilon)
\end{equation}
for any commuting states $\rho$ and $\sigma$ in $\,\S(\H_{A_1A_2})$ such that $\,\Tr H\rho_{A_1},\Tr H\sigma_{A_1}\leq E$ and $\frac{1}{2}\|\rho-\sigma\|_1\leq \varepsilon$.
Comparing the continuity bound (\ref{I-2-2}) with the universal continuity bounds for $I(A_1\!:\!A_2)$ under the energy constraint presented in \cite[Section 4.2.2]{QC}
we see an even greater gain in accuracy than in the first part of this example. Of course, this gain is due to the fact that we restrict attention to commuting states.
\end{example}

\section{Applications to characteristics of quantum and classical systems}

\subsection{Von Neumann entropy}

The first continuity bound for the von Neumann entropy in finite-dimensional quantum systems was obtained by  Fannes in \cite{Fannes}. This continuity bound and its optimized version obtained by Audenaert in \cite{Aud} are widely used in quantum information theory.  Audenaert's optimal continuity bound claims that
\begin{equation}\label{Aud}
  \vert S(\rho)-S(\sigma)\vert \leq \varepsilon\ln(d-1)+h_2(\varepsilon)
\end{equation}
for any states $\rho$ and $\sigma$ of a $d$-dimensional quantum system such that\break $\,\frac{1}{2}\|\rho-\sigma\|_1\leq \varepsilon\leq 1-1/d$. If
$\,\frac{1}{2}\|\rho-\sigma\|_1>1-1/d\,$ then the r.h.s. of (\ref{Aud}) must be replaced by $\ln d$.

In infinite dimensions two different continuity bounds for the von Neumann entropy under the energy constraints were obtained by Winter in \cite{W-CB}. The first of them
claims that
\begin{equation}\label{W-CB-1}
  \vert S(\rho)-S(\sigma)\vert \leq 2\varepsilon F_H(E/\varepsilon)+h_2(\varepsilon)
\end{equation}
for any states $\rho$ and $\sigma$ in $\S(\H)$ such that $\Tr H\rho,\Tr H\sigma\leq E$ and $\,\frac{1}{2}\|\rho-\sigma\|_1\leq \varepsilon$,
where $H$ is a positive operator on $\H$ satisfying conditions (\ref{H-cond})
and (\ref{star}) and $F_H$ is the function defined in (\ref{F-def}).\footnote{In \cite{W-CB} the function $F_{H}(E)$ is denoted by $S(\gamma(E))$.} The r.h.s. of (\ref{W-CB-1}) tends to zero as $\,\varepsilon\to0$ due to the equivalence of (\ref{H-cond}) and (\ref{H-cond-a}).

Continuity bound (\ref{W-CB-1}) is simple but not tight (because of factor 2 in the r.h.s.). On the contrary, Winter's second
continuity bound presented in Meta-Lemma 16 in \cite{W-CB} is not simple, but asymptotically tight for large energy $E$
provided that the function $F_H(E)$ has a logarithmic growth rate as $E\to+\infty$ (this property holds for Hamiltonians of many quantum systems, in particular, for multi-mode quantum oscillator \cite{B&D}).

Recently, Becker, Datta and  Jabbour obtained in \cite{BDJ} optimal continuity bound for the von Neumann entropy under the constraint induced by the number operator $\hat{N}=a^\dagger a\,$ of a one mode quantum oscillator. Namely, these authors proved that
\begin{equation}\label{BDJ-CB}
  \vert S(\rho)-S(\sigma)\vert \leq E h_2(\varepsilon/E)+h_2(\varepsilon)
\end{equation}
for any states $\rho$ and $\sigma$ in $\S(\H)$ such that $\Tr \hat{N}\rho,\Tr \hat{N}\sigma\leq E$ and $\,\frac{1}{2}\|\rho-\sigma\|_1\leq \varepsilon\leq E/(E+1)$.
They also showed that for any given $0<E<+\infty$ and $\varepsilon$ in $[0,E/(E+1)]$ there exist states
$\rho$ and $\sigma$ in $\S(\H)$ such that $\Tr \hat{N}\rho,\Tr \hat{N}\sigma\leq E$ and $\,\frac{1}{2}\|\rho-\sigma\|_1=\varepsilon$ for which an equality holds in (\ref{BDJ-CB}).

The results of numerical calculations presented in \cite{BDJ} show that continuity bound (\ref{BDJ-CB}) is much more accurate than
both of Winter's continuity bounds in the case $\,H=\hat{N}$.

By using the results of Section 3 one can obtain the following continuity bound for the von Neumann entropy under the constraint induced by any positive operator
$H$ satisfying conditions (\ref{H-cond}) and (\ref{star}):
\begin{equation}\label{Sh-CB}
   \vert S(\rho)-S(\sigma)\vert \leq \varepsilon F_H(E/\varepsilon)+g(\varepsilon)
\end{equation}
for any states $\rho$ and $\sigma$ in $\S(\H)$ such that $\Tr H\rho,\Tr H\sigma\leq E$ and $\,\frac{1}{2}\|\rho-\sigma\|_1\leq \varepsilon$, where $F_H$ and $g$ are the functions defined, respectively, in (\ref{F-def}) and (\ref{g-def}).

Comparing  continuity bound (\ref{Sh-CB}) with Winter's continuity bound (\ref{W-CB-1}), we see the disappearance of factor $2$ in r.h.s. what is paid for by replacing
the term $h_2(\varepsilon)$ with the term $g(\varepsilon)\geq h_2(\varepsilon)$. But keeping in mind that $\varepsilon F_H(E/\varepsilon)\gg g(\varepsilon)$ for
real values of $E$ and $\varepsilon$ and noting that  $g(\varepsilon)\sim h_2(\varepsilon)$ for small $\varepsilon$, we see that the
continuity bound (\ref{Sh-CB}) is more accurate in general. Moreover, continuity bound (\ref{Sh-CB}) is asymptotically tight for large energy $E$
for \emph{arbitrary} positive operator $H$ satisfying conditions (\ref{H-cond}) and (\ref{star}). This follows from the last claim of Proposition \ref{S-CB} below.

If $H=\hat{N}\doteq a^\dagger a\,$ then $F_H(E)=g(E)$ and continuity bound (\ref{Sh-CB}) becomes
\begin{equation}\label{Sh-CB+}
   \vert S(\rho)-S(\sigma)\vert \leq \varepsilon g(E/\varepsilon)+g(\varepsilon)=(E+\varepsilon)h_2\!\left(\frac{\varepsilon}{E+\varepsilon}\right)+(1+\varepsilon)h_2\!\left(\frac{\varepsilon}{1+\varepsilon}\right).
\end{equation}
We see that (\ref{Sh-CB+}) is less accurate but very close to the specialized optimal continuity bound (\ref{BDJ-CB}) for $E\gg\varepsilon$
and small $\varepsilon$. The difference between the right hand sides of (\ref{Sh-CB+}) and (\ref{BDJ-CB}) is equal to $2(g(\varepsilon)-h_2(\varepsilon))$ for $E=1$, and
quickly decreases to $g(\varepsilon)-h_2(\varepsilon)$ with increasing $E$. This and the last claim of Proposition \ref{S-CB} below give a reason to believe that the universal  continuity bound (\ref{Sh-CB}) is "close-to-optimal".\smallskip

Continuity bound (\ref{Sh-CB}) is a corollary of the following more general result.\smallskip

\begin{proposition}\label{S-CB} \emph{Let $H$ be a positive operator on $\H$  satisfying conditions (\ref{H-cond}) and (\ref{star}). If $\rho$ is a state in $\S(\H)$ such that $\Tr H\rho\leq E$ then
\begin{equation}\label{W-CB-2}
   S(\rho)-S(\sigma)\leq \varepsilon F_H((E-E_{H,\shs\varepsilon}(\rho))/\varepsilon)+g(\varepsilon)\leq \varepsilon F_H(E/\varepsilon)+g(\varepsilon)
\end{equation}
for any state $\sigma$ in $\S(\H)$ such that $\,\frac{1}{2}\|\rho-\sigma\|_1\leq \varepsilon$, where $E_{H,\shs\varepsilon}(\rho)\doteq \Tr H[\rho-\varepsilon I_{\H}]_+$ and the l.h.s. of (\ref{W-CB-2}) may be equal to $-\infty$.}\footnote{$\,[\rho-\varepsilon I_{\H}]_+$ is the positive part of the Hermitian operator $\,\rho-\varepsilon  I_{\H}$.}
\smallskip

\emph{The quantity $E_{H,\shs\varepsilon}(\rho)$ in (\ref{W-CB-2}) can be replaced by its easily-computable lower bound
\begin{equation*}
 E^*_{H,\shs\varepsilon}(\rho)\doteq \sum_{k=0}^{+\infty}E^H_k[\lambda^{\rho}_k-\varepsilon]_+\quad ([x]_+\doteq\max\{x,0\}),
\end{equation*}
where $\{\lambda^{\rho}_k\}_{k=0}^{+\infty}$ and $\{E^H_k\}_{k=0}^{+\infty}$ are the  sequences of eigenvalues of the state $\rho$ and the operator $H$ arranged, respectively, in the non-increasing and the non-decreasing orders.}

\smallskip

\emph{For each $E>0$ and $\varepsilon\in(0,1]$ there exist states $\rho$ and $\sigma$ in $\S(\H)$ such that}
$$
S(\rho)-S(\sigma)> \varepsilon F_H(E/\varepsilon),\quad \Tr H\rho,\Tr H\sigma\leq E\quad \textit{and} \quad \textstyle\frac{1}{2}\|\rho-\sigma\|_1\leq \varepsilon.
$$
\end{proposition}
\smallskip

\noindent\textbf{Note A:} In Proposition \ref{S-CB} we  impose no restrictions on the state $\sigma$ other than the requirement $\frac{1}{2}\|\rho-\sigma\|_1\leq \varepsilon$.\smallskip

\smallskip

\noindent\textbf{Note B:} The  quantity $E_{H,\shs\varepsilon}(\rho)$
monotonically tends to $\Tr H\rho\,$  as $\,\varepsilon\to 0$.
So, the first estimate in (\ref{W-CB-2}) with $E=\Tr H\rho$ may be essentially  sharper than the second one for small $\,\varepsilon$.
\smallskip

\begin{proof} Let $\rho$ and $\sigma$ be states in $\S(\H)$ such that $\Tr H\rho\leq E$ and $\,\frac{1}{2}\|\rho-\sigma\|_1\leq \varepsilon$.\smallskip

Let $\{\lambda^{\rho}_k\}_{k=0}^{+\infty}$ and $\{\lambda^{\sigma}_k\}_{k=0}^{+\infty}$ be the  sequences of eigenvalues of the states $\rho$ and $\sigma$ arranged in the non-increasing order and $\{\varphi_k\}_{k=0}^{+\infty}$ the basis in $\H$  such that
\begin{equation}\label{sp-st}
\rho=\sum_{k=0}^{+\infty} \lambda^{\rho}_k \vert \varphi_k\rangle\langle\varphi_k\vert.
\end{equation}
Consider the state
\begin{equation}\label{sp-st+}
\hat{\sigma}=\sum_{k=0}^{+\infty} \lambda^{\sigma}_k \vert \varphi_k\rangle\langle\varphi_k\vert.
\end{equation}
The Mirsky ineqiality (\ref{Mirsky-ineq+}) implies that
$$
\textstyle\frac{1}{2}\|\rho-\hat{\sigma}\|_1\leq\frac{1}{2}\|\rho-\sigma\|_1\leq \varepsilon.
$$
So, since the von Neumann entropy belongs to the class $L_1^{1}(1,1\vert \S(\H))$ and the states $\rho$ and $\hat{\sigma}$ commute, Corollary \ref{main-c}
with $\S_0=\S(\H)$ shows that
$$
S(\rho)-S(\sigma)=S(\rho)-S(\hat{\sigma})\leq \varepsilon F_H((E-\Tr H[\rho-\varepsilon I_{\H}]_+)/\varepsilon)+g(\varepsilon).
$$
The inequality $\,E^*_{H,\shs\varepsilon}(\rho)\leq E_{H,\shs\varepsilon}(\rho)\,$ follows from the Courant-Fisher theorem (see  Proposition 2.3. in \cite{BDJ}).

To prove the last claim one can take the states $\rho=\varepsilon\gamma_H(E/\varepsilon)+(1-\varepsilon)\vert \tau_0\rangle\langle\tau_0\vert$ and $\sigma=\vert \tau_0\rangle\langle\tau_0\vert$,
where $\gamma_H(E/\varepsilon)$ is the Gibbs state (\ref{Gibbs}) corresponding to the "energy" $E/\varepsilon$ and $\tau_0$ is the eigenvector of $H$ corresponding to the
minimal eigenvalue $E_0=0$.
\end{proof}

Proposition \ref{S-CB} implies the following refinement of continuity bound (\ref{Sh-CB}) in the case of states with different energies.\smallskip

\begin{corollary}\label{S-CB-c} \emph{If $H$ is a positive operator on $\H$ satisfying conditions (\ref{H-cond}) and (\ref{star}) then
\begin{equation*}
 -(\varepsilon F_H(E_{\sigma}/\varepsilon)+g(\varepsilon)) \leq S(\rho)-S(\sigma)\leq \varepsilon F_H(E_{\rho}/\varepsilon)+g(\varepsilon)
\end{equation*}
for any states $\rho$ and $\sigma$ in $\S(\H)$ such that $E_{\rho}\doteq\Tr H\rho<+\infty$, $E_{\sigma}\doteq\Tr H\sigma<+\infty$ and
$\,\frac{1}{2}\|\rho-\sigma\|_1\leq \varepsilon$.}
\end{corollary}\smallskip

By using Corollary \ref{main-1-c} one can prove the version of Proposition \ref{S-CB} in which the
condition $\Tr H\rho<+\infty$ is replaced by the condition $\rank\rho<+\infty$. It states that
\begin{equation*}
   S(\rho)-S(\sigma)\leq \varepsilon\ln(\rank\rho)+g(\varepsilon)
\end{equation*}
for any state $\sigma$ in $\S(\H)$ such that $\,\frac{1}{2}\|\rho-\sigma\|_1\leq \varepsilon$. But
by using Audenaert's optimal continuity bound (\ref{Aud}) one can prove  more sharp inequality.\smallskip

\begin{proposition}\label{S-CB+} \emph{Let $\rho$ be a finite rank state in $\S(\H)$.\footnote{We assume that $\dim\H=+\infty$.} Then
\begin{equation}\label{S-CB-1+}
   S(\rho)-S(\sigma)\leq \varepsilon\ln(\rank\rho-1)+h_2(\varepsilon)
\end{equation}
for any state $\sigma$ in $\S(\H)$ such that $\,\frac{1}{2}\|\rho-\sigma\|_1\leq \varepsilon\leq 1-1/\rank\rho$ (the l.h.s. of (\ref{S-CB-1+}) may be equal to $-\infty$).}
\end{proposition}

\begin{proof} By the arguments from the proof of Proposition \ref{S-CB} we may assume that
the states $\rho$ and $\sigma$ have representations (\ref{sp-st}) and (\ref{sp-st+}) correspondingly.

Let $n=\rank\rho<+\infty$. Consider the quantum channel
$$
\Phi(\varrho)=\sum_{k=0}^{+\infty}W_k\varrho W_k^*,\quad \varrho\in\S(\H),
$$
where $\,W_k=\sum_{i=0}^{n-1}\vert \varphi_i\rangle\langle\varphi_{i+nk}\vert\,$ is a partial isometry such that
$W_kW^*_k$ and $W^*_kW_k$ are the projectors on the linear spans of the sets $\,\{\varphi_0,...,\varphi_{n-1}\}\,$ and $\,\{\varphi_{nk},...,\varphi_{nk+n-1}\}\,$
correspondingly.

Let $\tilde{\sigma}=\Phi(\sigma)$. Both states $\rho$ and $\tilde{\sigma}$ are supported by the $n$-dimensional subspace
generated by the vectors $\varphi_0,...,\varphi_{n-1}$. Since $\rho=\Phi(\rho)$ by the construction, we have
$$
\|\rho-\tilde{\sigma}\|_1=\|\Phi(\rho)-\Phi(\sigma)\|_1\leq\|\rho-\sigma\|_1\leq2\varepsilon
$$
due to monotonicity of the trace norm under action of a channel. Thus, to derive the claim of Proposition \ref{S-CB+}
from  Audenaert's continuity bound (\ref{Aud}) it suffices to show that $S(\tilde{\sigma})\leq S(\sigma)$. This can be done by noting that
$$
\tilde{\sigma}=\sum_{k=0}^{+\infty}\lambda^{\sigma}_{kn}\vert \varphi_0\rangle\langle\varphi_0\vert +...+\sum_{k=0}^{+\infty}\lambda^{\sigma}_{kn+i}\vert \varphi_i\rangle\langle\varphi_i\vert +...
+\sum_{k=0}^{+\infty}\lambda^{\sigma}_{kn+n-1}\vert \varphi_{n-1}\rangle\langle\varphi_{n-1}\vert,
$$
and hence the state $\sigma$ is majorized by the state $\tilde{\sigma}$ in the sense of \cite[Section 13.5]{S-T-1}.
\end{proof}

By combining Propositions \ref{S-CB} and \ref{S-CB+} one can obtain the "mixed" continuity bound:
\begin{equation*}
-(\varepsilon\ln(d-1)+h_2(\varepsilon))\leq S(\rho)-S(\sigma)\leq \varepsilon F_H(E/\varepsilon)-g(\varepsilon)
\end{equation*}
for any states $\rho$ and $\sigma$ in $\S(\H)$ such that $\Tr H\rho\leq E$, $\rank\sigma \leq d\,$ and\break $\,\frac{1}{2}\|\rho-\sigma\|_1\leq \varepsilon$ provided that $\,\varepsilon\leq 1-1/d\,$  and  $H$ is a positive operator on $\H$ satisfying conditions (\ref{H-cond}) and (\ref{star}).

\subsection{Quantum conditional entropy of quantum-classical states}

In this subsection we apply the results of Section 3 to obtain continuity bounds and semi-continuity bounds for the (extended) quantum conditional entropy (QCE) defined in (\ref{ce-ext}) restricting attention to
quantum-classical (q-c) states of a bipartite system $AB$, i.e. states $\rho$ and $\sigma$ having the form
\begin{equation}\label{qc-states}
\rho=\sum_{k} p_k\, \rho_k\otimes \vert k\rangle\langle k\vert \quad \textrm{and}\quad \sigma=\sum_{k} q_k\, \sigma_k\otimes \vert k\rangle\langle k\vert,
\end{equation}
where $\{p_k,\rho_k\}$ and $\{q_k,\sigma_k\}$ are ensembles of states in $\S(\H_A)$ and $\{\vert k\rangle\}$ a fixed orthonormal basis in $\H_B$.
By using definition (\ref{ce-ext}) one can show (see the proof of Corollary 3 in \cite{Wilde-CB}) that
\begin{equation}\label{ce-rep}
S(A\vert B)_{\rho}=\sum_kp_kS(\rho_k)\leq+\infty\quad \textrm{and} \quad S(A\vert B)_{\sigma}=\sum_kq_kS(\sigma_k)\leq+\infty.
\end{equation}
The expressions in (\ref{ce-rep}) allow us to define the QCE on the set $\S_{\mathrm{qc}}$ of all q-c states
(including the q-c states $\rho$ with $\,S(\rho_A)=+\infty\,$ for which definition (\ref{ce-ext}) does not work).
The properties of the von Neumann entropy imply that the QCE defined in such a way on the convex set $\S_{\mathrm{qc}}$ satisfies
inequalities (\ref{LAA-1}) and (\ref{LAA-2}) with $a_f=0$ and $b_f=h_2$ with possible values $+\infty$ in both sides.

In finite dimensions the first continuity bound for the QCE was obtained by Alicki and Fannes in \cite{A&F}. Then
Winter optimized this continuity bound and obtained its specification for q-c states \cite{W-CB}. In 2019, by using the Alhejji-Smith  optimal  continuity bound for the Shannon conditional entropy (presented in \cite{A&G}) Wilde proved optimal continuity bound for the QCE restricted to q-c states \cite{Wilde-CB}.

The first continuity bound for the QCE in infinite-dimensional quantum system $AB$  under the energy-type constraint imposed on
subsystem $A$ was obtained by Winter \cite{W-CB}. Another
continuity bound for the QCE under the same constraint is presented in Proposition 5 in \cite[Section 4.1.4]{QC}.
Both continuity bounds are universal and cannot be improved by restricting attention to q-c states.

To explain our approach take any $\varepsilon>0$ and assume that  $\rho$ and $\sigma$ are q-c states  with representation (\ref{qc-states}) such that
$$
\|\rho-\sigma\|_1=\sum_k\|p_k\rho_k-q_k\sigma_k\|_1\leq 2\varepsilon.
$$
For each $k$ let $\{\varphi^k_i\}$ be the orthonormal basis in $\H_A$ such that
$\rho_k=\sum_{i} \lambda^{\rho_k}_i \vert \varphi^k_i\rangle\langle \varphi^k_i\vert$,
where $\{\lambda^{\rho_k}_i\}_i$ is a non-increasing sequence of eigenvalues of $\rho_k$.
Consider the q-c state
\begin{equation*}
\tilde{\sigma}=\sum_{k} q_k\tilde{\sigma}_k\otimes \vert k\rangle\langle k\vert ,\quad\tilde{\sigma}_k=\sum_{i} \lambda^{\sigma_k}_i \vert \varphi^k_i\rangle\langle \varphi^k_i\vert,
\end{equation*}
where $\{\lambda^{\sigma_k}_i\}_i$ is the non-increasing sequence of eigenvalues of $\sigma_k$.
By Mirsky inequality (\ref{Mirsky-ineq+}) we have
\begin{equation}\label{ce-one}
\|\rho-\tilde{\sigma}\|_1=\sum_k\|p_k\rho_k-q_k\tilde{\sigma}_k\|_1\leq\sum_k\|p_k\rho_k-q_k\sigma_k\|_1\leq 2\varepsilon.
\end{equation}
Note also that the representation (\ref{ce-rep}) implies that
\begin{equation}\label{ce-two}
S(A\vert B)_{\sigma}=\sum_kq_kS(\sigma_k)=\sum_kq_kS(\tilde{\sigma}_k)=S(A\vert B)_{\tilde{\sigma}}.
\end{equation}

By the remark after (\ref{ce-rep}) and due to upper bound (\ref{ce-ub}) the (extended) QCE $S(A\vert B)$ is a nonnegative function on
the convex set $\S_{\mathrm{qc}}$ of all q-c states with representation (\ref{qc-states}) belonging to the class $L^1_2(1,1\vert \S_{\mathrm{qc}})$.
So, since the set $\S_{\mathrm{qc}}$ possesses property (\ref{S-prop++}) and the states $\rho$ and $\tilde{\sigma}$ commute by the construction, by applying
Corollaries \ref{main-1-c}B and \ref{main-c}B in Section 3.2 with $\S_0=\S_{\mathrm{qc}}$ and taking  (\ref{ce-one}) and (\ref{ce-two}) into account
we obtain the following\smallskip\pagebreak

\begin{proposition}\label{qce-qc-1} \emph{Let $AB$ be an infinite-dimensional bipartite quantum system.}\smallskip

\emph{\noindent\emph{ A)} If $\rho$ is a q-c state in $\S(\H_{AB})$ such that $\rank\rho_A$ is finite then
\begin{equation}\label{qce-qc-1+}
   S(A\vert B)_{\rho}-S(A\vert B)_{\sigma}\leq \varepsilon\ln(\rank\rho_A)+g(\varepsilon)
\end{equation}
for any q-c state $\sigma$ in $\S(\H_{AB})$ such that $\,\frac{1}{2}\|\rho-\sigma\|_1\leq \varepsilon$ (the l.h.s. of (\ref{qce-qc-1+}) may be equal to $-\infty$).}\smallskip

\emph{\noindent\emph{ B)} Let $H$ be a positive operator on $\H_A$  satisfying conditions (\ref{H-cond}) and (\ref{star}). If $\rho$ is a q-c state in $\S(\H_{AB})$  such that $\Tr H\rho_A\leq E$ then
\begin{equation}\label{qce-qc-2}
   S(A\vert B)_{\rho}-S(A\vert B)_{\sigma}\leq \varepsilon F_H((E-E_{H,\shs\varepsilon}(\rho))/\varepsilon)+g(\varepsilon)\leq \varepsilon F_H(E/\varepsilon)+g(\varepsilon)
\end{equation}
for any q-c state $\sigma$ in $\S(\H_{AB})$ such that $\,\frac{1}{2}\|\rho-\sigma\|_1\leq \varepsilon$, where
\begin{equation*}
E_{H,\shs\varepsilon}(\rho)\doteq \sum_{k}\Tr H[p_k\rho_k-\varepsilon I_{A}]_+,
\end{equation*}
$\,[p_k\rho_k-\varepsilon  I_{A}]_+$ is the positive part of the Hermitian operator $\,p_k\rho_k-\varepsilon  I_{A}$ and the l.h.s. of (\ref{qce-qc-2}) may be equal to $-\infty$.}\footnote{Throughout this subsection we assume that  $\rho$ and $\sigma$ are q-c states
having representation (\ref{qc-states}).}\smallskip

\emph{For each $E>0$ and $\varepsilon\in(0,1]$ there exist q-c states $\rho$ and $\sigma$ in $\S(\H_{AB})$ such that}
$$
S(A\vert B)_{\rho}-S(A\vert B)_{\sigma}>\varepsilon F_H(E/\varepsilon),\quad \Tr H\rho_A,\Tr H\sigma_A\leq E\quad \textit{and} \quad \textstyle\frac{1}{2}\|\rho-\sigma\|_1\leq \varepsilon.
$$
\end{proposition}\smallskip

The last claim of Proposition \ref{qce-qc-1} can be proved by taking
the q-c states $\rho\otimes\vert0\rangle\langle0\vert$ and $\sigma\otimes\vert0\rangle\langle0\vert$, where  $\rho$ and $\sigma$ are the states
from the last claim of Proposition \ref{S-CB}.\smallskip


\noindent\textbf{Note:} The  quantity $E_{H,\shs\varepsilon}(\rho)$
monotonically tends to $\Tr H\rho_A\,$  as $\,\varepsilon\to 0$.
So, the first estimate in (\ref{qce-qc-2})  with $E=\Tr H\rho_A$ may be essentially  sharper than the second one for small $\,\varepsilon$.

It follows from Proposition \ref{qce-qc-1}A that
\begin{equation}\label{qce-qc-1++}
   -(\varepsilon\ln(\rank\sigma_A)+g(\varepsilon))\leq S(A\vert B)_{\rho}-S(A\vert B)_{\sigma}\leq \varepsilon\ln(\rank\rho_A)+g(\varepsilon)
\end{equation}
for any q-c states $\rho$ and $\sigma$ in $\S(\H_{AB})$ such that $\,\frac{1}{2}\|\rho-\sigma\|_1\leq \varepsilon$.

Since the dimension of the subspace generated by  the supports of  $\rho_A$ and $\sigma_A$ may be
close to $\,\rank\rho_A+\rank\sigma_A$, in some cases the continuity bound  (\ref{qce-qc-1++}) may be sharper than
Wilde's optimal continuity bound for QCE of q-c states proposed in \cite{Wilde-CB}.

Proposition \ref{qce-qc-1}B implies that
\begin{equation}\label{qce-qc-2++}
\vert S(A\vert B)_{\rho}-S(A\vert B)_{\sigma}\vert \leq \varepsilon F_H(E/\varepsilon)+g(\varepsilon)
\end{equation}
for any q-c states $\rho$ and $\sigma$ in $\S(\H_{AB})$ such that $\Tr H\rho_A, \Tr H\sigma_A\leq E$  and $\,\frac{1}{2}\|\rho-\sigma\|_1\leq \varepsilon$, where $H$ is a positive operator on $\H_A$ satisfying conditions (\ref{H-cond}) and (\ref{star}). It is easy to see that continuity bound (\ref{qce-qc-2++}) is \emph{essentially sharper} than the
existing continuity bounds for the QCE under the energy-type constraints mentioned before. The last claim of  Proposition \ref{qce-qc-1}
shows that continuity bound  (\ref{qce-qc-2++}) is \emph{close to optimal}. It can be treated as
a generalization of continuity bound (\ref{Sh-CB}) for the von Neumann entropy: inequality (\ref{qce-qc-2++}) with the q-c states $\rho\otimes\vert 0\rangle\langle0\vert $ and $\sigma\otimes\vert 0\rangle\langle0\vert $ turns into inequality (\ref{Sh-CB}).

\subsection{Entanglement of formation}

The entanglement of
formation (EoF) is one of the basic entanglement measures in bipartite quantum systems \cite{Bennett,4H,P&V}. In a finite-dimensional bipartite system $AB$ the EoF is defined as the convex roof extension to the set $\S(\H_{AB})$ of the function $\omega\mapsto S(\omega_{A})$ on the set $\mathrm{ext}\shs\S(\H_{AB})$ of pure states in $\S(\H_{AB})$ , i.e.
\begin{equation*}
  E_{F}(\omega)=\inf_{\sum_kp_k\omega_k=\omega}\sum_k p_kS([\omega_k]_{A}),
\end{equation*}
where the infimum is over all finite ensembles $\{p_k, \omega_k\}$ of pure states in $\S(\H_{AB})$ with the average state $\omega$ \cite{Bennett}.
In this case $E_F$ is a uniformly continuous function on $\S(\H_{AB})$  possessing basic properties of an entanglement measure \cite{4H,P&V,Wilde-new}.

The first continuity bound for the EoF in a finite-dimensional bipartite system $AB$ was obtained by Nielsen \cite{Nielsen}. Then it was improved by  Winter who used his continuity bound for the QCE of q-c states and updated Nielsen's arguments \cite{W-CB}. Finally, Wilde embedded his optimal continuity bound for the QCE  of q-c states into Nielsen-Winter construction to obtain the most accurate continuity bound for the EoF to date \cite{Wilde-CB}.

If both systems $A$ and $B$ are infinite-dimensional then  there are two versions $E_F^d$ and $E_F^c$ of the EoF defined, respectively, by means of discrete and continuous convex roof extensions
\begin{equation}\label{E_F-def-d}
E_F^d(\omega)=\!\inf_{\sum_k\!p_k\omega_k=\omega}\sum_kp_kS([\omega_k]_A),\quad
\end{equation}
\begin{equation}\label{E_F-def-c}
E_F^c(\omega)=\!\inf_{\int\omega'\mu(d\omega')=\omega}\int\! S(\omega'_A)\mu(d\omega'),
\end{equation}
where the  infimum in (\ref{E_F-def-d}) is over all countable ensembles $\{p_k, \omega_k\}$ of pure states in $\S(\H_{AB})$ with the average state $\omega$ and the  infimum in (\ref{E_F-def-c}) is over all Borel probability measures on the set of pure states in $\S(\H_{AB})$ with the barycenter $\omega$ (the last infimum is always attained) \cite[Section 5]{EM}. It follows from the definitions that $E_F^d(\omega)\geq E_F^c(\omega)$ for any state $\omega\in\S(\H_{AB})$. In \cite{EM} it is shown that the function $E_F^c$ has better properties (as an entanglement measure), in particular, it is
lower semicontinuous on $\S(\H_{AB})$ and is equal to zero at any separable state in $\S(\H_{AB})$ (as  far as I know, the equality $E_F^d(\omega)=0$  has not yet been proven if $\omega$ is a countably-non-decomposable separable state such that $S(\omega_{A})=S(\omega_{B})=+\infty$  \cite[Remark 6]{EM}).

In \cite{EM} it is shown that $E_F^d(\omega)=E_F^c(\omega)$ for any state $\omega$ in $\S(\H_{AB})$ such that $\min\{S(\omega_{A}),S(\omega_{B})\}<+\infty$.  The conjecture of coincidence of $E_F^d$ and $E_F^c$ on $\S(\H_{AB})$ is an interesting open question.

Different continuity bounds for the functions $E_F^d$ and $E_F^c$ under the energy-type constraints are obtained by using the Nielsen-Winter technique and different continuity bounds for the QCE.  They are described
in Section 4.4.2 in \cite{QC}.

By using the semi-continuity bounds for the QCE of q-c states presented in Proposition \ref{qce-qc-1} and the Nielsen-Winter technique
one can obtain semi-continuity bounds for the functions $E_F^d$ and $E_F^c$.\pagebreak

\begin{proposition}\label{EF-CB} \emph{Let $AB$ be an infinite-dimensional bipartite quantum system.}\smallskip

\emph{\noindent \emph{A)} If $\rho$ is a state in $\S(\H_{AB})$ such that $\rank\rho_A$ is finite then
\begin{equation}\label{EF-CB-A}
   E^*_F(\rho)-E^*_F(\sigma)\leq \delta\ln(\rank\rho_A)+g(\delta),\quad E^*_F=E_F^d,E_F^c,
\end{equation}
for any state $\sigma$ in $\S(\H_{AB})$ such that $\,\frac{1}{2}\|\rho-\sigma\|_1\leq \varepsilon\leq1$,  where $\delta=\sqrt{\varepsilon(2-\varepsilon)}$ and  the l.h.s. of (\ref{EF-CB-A}) may be equal to $-\infty$.}\smallskip

\emph{\noindent \emph{B)} If $\rho$ is a state in $\S(\H_{AB})$ such that $\Tr H\rho_A\leq E$, where $H$ is a positive operator on $\H_A$  satisfying conditions (\ref{H-cond}) and (\ref{star}), then
\begin{equation}\label{EF-CB-B}
   E^*_F(\rho)-E^*_F(\sigma)\leq \delta F_H(E/\delta)+g(\delta),\quad E^*_F=E_F^d,E_F^c,
\end{equation}
for any state $\sigma$ in $\S(\H_{AB})$ such that $\,\frac{1}{2}\|\rho-\sigma\|_1\leq \varepsilon\leq1$, where $\,\delta=\sqrt{\varepsilon(2-\varepsilon)}$ and  the l.h.s. of  (\ref{EF-CB-B}) may be equal to $-\infty$.}
\end{proposition}\smallskip


\noindent\textbf{Note A:} The conditions imposed in both parts of Proposition \ref{EF-CB} on the state
$\rho$ imply that $E_F^c(\rho)=E_F^d(\rho)<+\infty$ (since they imply that $S(\rho_A)<+\infty$). \smallskip

\begin{proof} Both claims of the proposition for  $E^*_F=E_F^d$ are obtained from the semi-continuity bounds for the QCE of q-c states presented in Proposition \ref{qce-qc-1} by using  the Nielsen-Winter technique (see the proof of Corollary 4 in \cite{W-CB}). The only difference consists in the necessity to use an $\epsilon$-optimal decomposition of the state $\sigma$, because an optimal decomposition may not exist in the infinite-dimensional case.

The claims for  $E^*_F=E_F^c$ can be proved by a simple approximation.  We may assume that $E_F^c(\sigma)<+\infty$  (since otherwise (\ref{EF-CB-A}) and (\ref{EF-CB-B}) obviously hold). We may also assume that $\,\delta=\sqrt{\varepsilon(2-\varepsilon)}\neq0$, where $\varepsilon=\frac{1}{2}\|\rho-\sigma\|_1$.

Let $\{P_n\}$ be any  sequence of finite-rank projectors in $\B(\H_A)$ strongly converging to the unit operator $I_A$. Then the sequence
of states $\sigma_n\doteq c_n^{-1}(P_n\otimes I_B)\sigma(P_n\otimes I_B)$, $c_n= \Tr P_n\sigma_A$, (well defined for all $n$ large enough) converges to the state $\sigma$.

By the monotonicity of $E_F^c$ under selective LOCC operations
(which follows from part A-2 of Theorem 2 in \cite{EM}) we have $c_nE_F^c(\sigma_n)\leq E_F^c(\sigma)$ for all $n$. So, since $c_n\to 1$ as $n\to+\infty$, the
lower semicontinuity of $E_F^c$ (mentioned after (\ref{E_F-def-c})) implies that
\begin{equation}\label{rl-0211}
\lim_{n\to+\infty} E_F^c(\sigma_n)=E_F^c(\sigma)<+\infty.
\end{equation}
Since $E_F^c(\sigma_n)=E_F^d(\sigma_n)$ for all $n$ (because $S([\sigma_n]_A)<+\infty$) and $E_F^c(\rho)=E_F^d(\rho)$ (by Note A), the proved claims of the proposition show that
the inequalities (\ref{EF-CB-A}) and (\ref{EF-CB-B}) with $\,E^*_F=E_F^c\,$ hold provided that $\sigma$ and $\delta$ are replaced, respectively, by $\sigma_n$
and $\,\delta_n=\sqrt{\varepsilon_n(2-\varepsilon_n)}$, $\varepsilon_n=\frac{1}{2}\|\rho-\sigma_n\|_1$, for all $n$. Since  $\,\delta_n$ tends to $\,\delta>0$ as $n\to+\infty$, by taking the limit as $n\to+\infty$  one can prove, due to  (\ref{rl-0211}), the validity of inequalities (\ref{EF-CB-A}) and (\ref{EF-CB-B}) with $\,E^*_F=E_F^c\,$
(as $F_H$ is a continuous function on $(0,+\infty)$ by the condition (\ref{star})).
\end{proof}

Proposition \ref{EF-CB}A implies that
\begin{equation}\label{EF-CB-1}
   -(\delta\ln(\rank\sigma_A)+g(\delta))\leq E_F(\rho)-E_F(\sigma)\leq \delta\ln(\rank\rho_X)+g(\delta)
\end{equation}
for any states $\rho$ and $\sigma$ in $\S(\H_{AB})$ such that  $\,\frac{1}{2}\|\rho-\sigma\|_1\leq \varepsilon\leq1$ and $\rank\rho_X,\rank\sigma_A<+\infty$, where $\delta=\sqrt{\varepsilon(2-\varepsilon)}$,  $X$ is either $A$ or $B$ and $E_F(\omega)\doteq E_F^d(\omega)=E_F^c(\omega)$, $\omega=\rho,\sigma$ (the equalities $E_F^d(\omega)=E_F^c(\omega)$, $\omega=\rho,\sigma$, follow from the conditions $\rank\rho_X,\rank\sigma_A<+\infty$).

Continuity bound  (\ref{EF-CB-1}) with $X=A$ may be sharper than the continuity bound for the EoF obtained by Wilde in \cite{Wilde-CB}
in the case when the dimension of the subspace generated by  the supports of states $\rho_A$ and $\sigma_A$ is
close to $\,\rank\rho_A+\rank\sigma_A$. Continuity bound  (\ref{EF-CB-1}) with $X=B$  has no analogues and gives a way to obtain
more accurate estimate for the quantity $E_F(\rho)-E_F(\sigma)$.

Let $H$ be a positive operator on $\H_A$ satisfying conditions (\ref{H-cond}) and (\ref{star}). Proposition \ref{EF-CB}B implies that
\begin{equation}\label{EF-CB-2}
\vert E_F(\rho)-E_F(\sigma)\vert \leq \delta F_H(E/\delta)+g(\delta)
\end{equation}
for any states $\rho$ and $\sigma$ in $\S(\H_{AB})$ such that $\Tr H\rho_A, \Tr H\sigma_A\leq E<+\infty$  and\break
$\,\frac{1}{2}\|\rho-\sigma\|_1\leq \varepsilon\leq1$,  where $\delta=\sqrt{\varepsilon(2-\varepsilon)}$
and $E_F(\omega)\doteq E_F^d(\omega)=E_F^c(\omega)$, $\omega=\rho,\sigma$ (the equalities $E_F^d(\omega)=E_F^c(\omega)$, $\omega=\rho,\sigma$, follow from the conditions $\Tr H\rho_A, \Tr H\sigma_A\leq E$, since they imply that $S(\rho_A), S(\sigma_A)<+\infty$ \cite{W}). Continuity bound (\ref{EF-CB-2}) is \emph{essentially sharper} than all the
existing continuity bounds for the  EoF under the energy-type constraint (described in Section 4.4.2 in \cite{QC}).\smallskip


\begin{remark}\label{fidelity}
In all the continuity bounds and the semi-continuity bounds for the EoF presented
in this subsection  the condition $\,\frac{1}{2}\|\rho-\sigma\|_1\leq \varepsilon$, where
$\varepsilon$ is such that $\,\delta=\sqrt{\varepsilon(2-\varepsilon)}$, can be replaced by the
condition  $\,F(\rho,\sigma)\geq 1-\delta^2$, where  $F(\rho,\sigma)\doteq\|\sqrt{\rho}\sqrt{\sigma}\|_1^2$
is the fidelity of $\rho$ and $\sigma$ \cite{H-SCI,Wilde}. This follows from the  Nielsen-Winter arguments used
in their proofs. The use of fidelity as a measure of closeness of the
states $\rho$ and $\sigma$ makes these continuity bounds essentially sharper in some cases.
\end{remark}

\subsection{Characteristics of discrete random variables}

Assume that $\{\tau^1_i\}_{i\in\N}$,...,$\{\tau^n_i\}_{i\in\N}\,$ are given orthonormal bases in separable Hilbert spaces $\H_{A_1}$,...,$\H_{A_n}$
correspondingly. Then the  set $\S_\tau$ of states in $\S(\H_{A_1...A_n})$ diagonizable in the basis
\begin{equation}\label{c-basis}
 \{\tau^1_{i_1}\otimes\cdots\otimes\tau^n_{i_n}\}_{(i_1,..,i_n)\in\N^n}
\end{equation}
can be identified with the set $\mathfrak{P}_n$
of $n$-variate probability distributions $\{p_{i_1,..,i_n}\}_{(i_1,..,i_n)\in\N^n}$.

Any $n$-variate probability distribution $\bar{p}=\{p_{i_1,..,i_n}\}_{(i_1,..,i_n)\in\N^n}$ can be treated as a
joint  distribution of some discrete random variables $X_1$,...,$X_n$. So, a value of some entropic characteristic of $n$-partite quantum system $A_1...A_n$  at
the state $\omega\in\S_\tau$ identified with the distribution $\bar{p}$ coincides with the value of the corresponding classical characteristic of the random variables $X_1$,...,$X_n$. For example, the quantum mutual information $I(A_1\!:\!A_2)$ of a bipartite state $\omega=\sum_{i,j}p_{i,j}\vert \tau^1_i\otimes\tau^2_j\rangle\langle\tau^1_i\otimes\tau^2_j\vert $
is equal to the classical mutual information $I(X_1\!:\!X_2)$ of random variables $X_1$,$X_2$ having joint distribution $\{p_{i,j}\}$.

Thus, since the set $\S_\tau$ consists of commuting states, one can apply Corollaries \ref{main-1-c} and \ref{main-c} in Section 3.2 to obtain continuity bounds for entropic  characteristics of classical discrete random variables under constrains of different types.

To formulate the main results of this section we have to introduce  "classical" analogues of the classes
$L_n^m(C,D\vert \S_0)$ described in Section 3.2.  Denote by
$T_n^m(C,D\vert \shs\PP_0)$ the class of all functions on a convex subset $\PP_0$ of $\PP_n$ taking values in $(-\infty,+\infty]$ that
satisfy the inequalities
\begin{equation*}
  f(\lambda\bar{p}+(1-\lambda)\bar{q})\geq \lambda f(\bar{p})+(1-\lambda)f(\bar{q})-d_f^-h_2(\lambda)
\end{equation*}
and
\begin{equation*}
  f(\lambda\bar{p}+(1-\lambda)\bar{q})\leq \lambda f(\bar{p})+(1-\lambda)f(\bar{q})+d_f^+h_2(\lambda)
\end{equation*}
for all distributions  $\bar{p}$ and $\bar{q}$ in $\PP_0$ and any $\lambda\in[0,1]$ with possible infinite values in both sides, where $d^-_f$ and $d^+_f$ are nonnegative numbers such that $d^-_f+d^+_f=D$, and the double inequality
\begin{equation}\label{Cm-c}
-c^-_f C_m(\bar{p})\leq f(\bar{p})\leq c^+_f C_m(\bar{p}),\quad C_m(\bar{p})=\sum_{k=1}^m H(\bar{p}_{k}),
\end{equation}
for any distribution $\bar{p}$  in $\PP_0$ with possible infinite values in all sides, where $\bar{p}_k$ denotes the marginal distribution of $\bar{p}$ corresponding to the $k$-th component, i.e. $\,[\bar{p}_k]_i=\sum_{(i_1,..,i_{n})\setminus\{i_k\}} p_{i_1,..,i_{k-1},i,i_{k+1},..,i_n}$, $H(\cdot)$
is the Shannon entropy and $c^-_f$ and $c^+_f$ are nonnegative numbers such that $c^-_f+c^+_f=C$.\footnote{The definitions of all the notions from the classical
information theory  used in this section can be found in \cite{C&T}.}

We will denote by $T_n^m(C,D)$ the class $T_n^m(C,D\vert \shs\PP^m_n)$, where
$$
\mathfrak{P}^m_n=\left\{\bar{p}\in\mathfrak{P}_n\,\vert \,H(\bar{p}_k)<+\infty,\,k=\overline{1,m}\right\}
$$
is the maximal set on which all the functions satisfying (\ref{Cm-c}) are finite.

Within this notation the Shannon entropy belongs to the class $T^1_1(1,1\vert \shs\PP_1)$ (due to its concavity and the classical version of inequality (\ref{w-k-ineq})).
The Shannon conditional entropy $H(X_1\vert X_2)$ (also called equivocation) extended to the set $\mathfrak{P}^1_2$ by the classical version of formula (\ref{ce-ext}) belongs to the class $T^1_2(1,1)\doteq T^1_2(1,1\vert\shs\mathfrak{P}^1_2)$. This follows from its nonnegativity, concavity and the classical versions of inequalities  (\ref{ce-ub}) and (\ref{ce-LAA-2}).

The mutual information $I(X_1\!:\!...\!:\!X_n)$ (also called total correlation \cite{Wat}) defined on the set $\mathfrak{P}_n$ by the classical version of formula (\ref{mi-d}) belongs to the classes  $T^{n-1}_n(1,n\vert \shs\PP_n)$ and $T^{n}_n(1-1/n,n\vert \shs\PP_n)$. This can be shown by using
its nonnegativity, the inequalities\footnote{These inequalities are the classical versions of the inequalities in \cite[formula (10)]{CBM}.}
$$
I(X_1\!:\!...\!:\!X_n)_{\lambda\bar{p}+(1-\lambda)\bar{q}}\geq \lambda I(X_1\!:\!...\!:\!X_n)_{\bar{p}}+(1-\lambda)I(X_1\!:\!...\!:\!X_n)_{\bar{q}}-h_2(\lambda),
$$
$$
I(X_1\!:\!...\!:\!X_n)_{\lambda\bar{p}+(1-\lambda)\bar{q}}\leq \lambda I(X_1\!:\!...\!:\!X_n)_{\bar{p}}+(1-\lambda)I(X_1\!:\!...\!:\!X_n)_{\bar{q}}+(n-1)h_2(\lambda)
$$
valid for all distributions  $\bar{p}$ and $\bar{q}$ in $\mathfrak{P}_n$ and any $\lambda\in[0,1]$ with possible value $+\infty$ in both sides, and the upper bounds
$$
I(X_1\!:\!...\!:\!X_n)_{\bar{p}}\leq \sum_{k=1}^{n-1}H(\bar{p}_k),\quad I(X_1\!:\!...\!:\!X_n)_{\bar{p}}\leq \frac{n-1}{n}\sum_{k=1}^{n}H(\bar{p}_k)
$$
(the first upper bound follows from the classical version of the representation in \cite[formula (9)]{CBM} with trivial system $C$ and upper bound (\ref{MI-UB}) along with the remark after it, the second one is obtained from the first by simple symmetry arguments).

The one-to-one correspondence between  the set $\PP_n$ of all $n$-variate probability distributions and the set $\S_{\tau}$ of quantum states in $\S(\H_{A_1...A_n})$ diagonizable in basis
(\ref{c-basis}) allows us to identify the class $T^m_n(C,D\vert \shs\PP_0)$ with the class $L^m_n(C,D\vert \shs\PP_0^{\tau})$, where $\PP_0^{\tau}$ is the subset of $\S_{\tau}$
corresponding to the subset $\PP_0$. So, we may  directly apply Corollaries \ref{main-1-c} and \ref{main-c} to obtain continuity bounds
for functions from the classes $T^m_n(C,D\vert \shs\PP_0)$.

We will use the total variation distance  (\ref{TVD-def}) between $n$-variate probability distributions $\bar{p}=\{p_{i_1,..,i_n}\}$ and $\bar{q}=\{q_{i_1,..,i_n}\}$
(considered as probability measures on $\mathbb{N}^{n}$). In this case we have
$$
\mathrm{TV}(\bar{p},\bar{q})=\textstyle\frac{1}{2}\displaystyle\sum_{i_1,..,i_n}\vert p_{i_1,..,i_n}-q_{i_1,..,i_n}\vert =\textstyle\frac{1}{2}\|\rho-\sigma\|_1,
$$
where $\rho$ and $\sigma$ are the states in $\S_{\tau}$ corresponding to the distributions $\bar{p}$ and $\bar{q}$.\smallskip

Corollary \ref{main-1-c} Section 3.2. with $\S_0=\PP_0^{\tau}$ implies the following\smallskip

\begin{proposition}\label{main-1-p} \emph{Let  $\,\PP_0$ be a convex subset of $\,\PP_n$ with the property
\begin{equation}\label{S-prop+++}
  \Delta^\pm(\bar{p},\bar{q})\in\PP_0\;\textrm{ for any }\,\bar{p}\,\textrm{ and }\,\bar{q}\textrm{ in }\,\PP_0\,\textrm{ such that }\, \bar{p}\neq\bar{q},
\end{equation}
where $\Delta^+(\bar{p},\bar{q})$ and $\Delta^-(\bar{p},\bar{q})$ are the $n$-variate probability distributions with the entries $c[p_{i_1,..,i_n}-q_{i_1,..,i_n}]_+$
and $\,c[q_{i_1,..,i_n}-p_{i_1,..,i_n}]_+$, $c=1/\mathrm{TV}(\bar{p},\bar{q})$\;  ($\,[x]_+=\max\{x,0\}$).}\smallskip

\emph{\noindent\emph{A)} If $f$ is a  function  in $T_n^{m}(C,D\vert \shs\PP_0)$  then
\begin{equation}\label{main+1-p}
    \vert f(\bar{p})-f(\bar{q})\vert \leq C\varepsilon \ln d_m(\bar{p},\bar{q})+Dg(\varepsilon)
\end{equation}
for any distributions $\bar{p}$ and $\bar{q}$ in $\,\PP_0$ such that $\,d_m(\bar{p},\bar{q})\doteq\max\left\{\prod_{k=1}^{m}\vert \bar{p}_k\vert , \prod_{k=1}^{m}\vert \bar{q}_k\vert \right\}$ is finite and $\mathrm{TV}(\bar{p},\bar{q})\leq \varepsilon$, where $\vert \bar{p}_k\vert $ and $\vert \bar{q}_k\vert $ are the numbers of nonzero entries of the $1$-variate marginal distributions $\bar{p}_k$ and $\bar{q}_k$ corresponding to the $k$-th component.}\smallskip

\emph{\noindent\emph{B)} If $f$ is a nonnegative function in $T_n^{m}(C,D\vert \shs\PP_0)$ and $\bar{p}$ is a distribution in $\,\PP_0$ such that $\,d_m(\bar{p})\doteq\prod_{k=1}^{m}\vert \bar{p}_k\vert $ is finite then
\begin{equation}\label{main++1-p}
    f(\bar{p})-f(\bar{q})\leq C\varepsilon \ln d_m(\bar{p})+Dg(\varepsilon)
\end{equation}
for any distribution  $\bar{q}$ in $\,\PP_0$ such that $\mathrm{TV}(\bar{p},\bar{q})\leq \varepsilon$ (the l.h.s. of (\ref{main++1-p}) may be equal to $-\infty$).}
\end{proposition}\smallskip

\begin{remark}\label{main-1-p-r}
It is easy to see that property (\ref{S-prop+++})
holds for the convex sets $\PP_n$ and $\PP_n^m$.
\end{remark}\smallskip

\noindent\textbf{Note A:} In  Proposition \ref{main-1-p}B  we  impose no restrictions on the distribution $\bar{q}$ other than the requirement $\mathrm{TV}(\bar{p},\bar{q})\leq \varepsilon$.

\smallskip

The "energy-constrained" version of Proposition \ref{main-1-p}  will be obtained for characteristics
belonging to the classes $T_n^1(C,D\vert \shs\PP_0)$ (for simplicity).

Let $\SC=\{E_i\}_{i=1}^{+\infty}$ be a nondecreasing sequence of nonnegative numbers such that
\begin{equation}\label{Z-cond}
\sum_{i=1}^{+\infty}e^{-\beta E_i}<+\infty\quad \forall\beta>0.
\end{equation}
Consider the function
\begin{equation}\label{F-Z}
\begin{array}{c}
\displaystyle F_\SC(E)=\sup\left\{H(\{p_i\})\,\left\vert\,\{p_i\}\in \mathfrak{P}_1,\; \sum_{i=1}^{+\infty} E_ip_i\leq E\right.\right\}\\
\displaystyle=\beta(E)E+\ln\sum_{i=1}^{+\infty} e^{-\beta(E)E_i},
\end{array}
\end{equation}
where $\beta(E)$ is defined by the equation $\sum_{i=1}^{+\infty} E_ie^{-\beta E_i}=E\sum_{i=1}^{+\infty} e^{-\beta E_i}$ \cite{W},\cite[Prop.1]{EC}.\smallskip

It is easy to see that the function $F_\SC(E)$ coincides with the function $F_H(E)$ defined in (\ref{F-def})
provided that $H$ is a positive operator with the spectrum $\SC$. If $\SC=\{0,1,2,...\}$ then $F_\SC(E)=g(E)$ -- the function defined in (\ref{g-def}). So,
Proposition 1 in \cite{EC} shows that condition (\ref{Z-cond}) is equivalent to the property
\begin{equation}\label{Z-cond+}
F_\SC(E)=o(E)\quad \textrm{as}\quad E\to+\infty.
\end{equation}

Thus, Corollary \ref{main-c} in Section 3.2 with $\S_0=\PP_0^{\tau}$ implies the following\smallskip

\begin{proposition}\label{main-p} \emph{Let  $\,\PP_0$ be a convex subset of $\,\PP_n$ with the property (\ref{S-prop+++}).
Let $\mathcal{S}=\{E_i\}_{i=1}^{+\infty}$ be a nondecreasing sequence of nonnegative numbers such that $E_1=0$
satisfying condition (\ref{Z-cond})
and $F_{\SC}$ the function defined in (\ref{F-Z}).}\smallskip

\emph{\noindent\emph{A)} If $f$ is a function in $T_n^{1}(C,D\vert \shs\PP_0)$ then
\begin{equation}\label{main+p}
    \vert f(\bar{p})-f(\bar{q})\vert \leq C\varepsilon F_{\SC}(E/\varepsilon)+Dg(\varepsilon)
\end{equation}
for any  distributions $\bar{p}$ and  $\bar{q}$ in $\,\mathfrak{P}_0$ such that $\,\sum_{i=1}^{+\infty} E_i[\bar{p}_1]_i,\,\sum_{i=1}^{+\infty} E_i[\bar{q}_1]_i\leq E\,$
and $\mathrm{TV}(\bar{p},\bar{q})\leq \varepsilon$, where $[\bar{p}_1]_i$ denotes the $i$-th entry of the marginal distribution $\bar{p}_1$.}\smallskip

\emph{\noindent\emph{B)} If $f$ is a nonnegative function in $T_n^{1}(C,D\vert \shs\PP_0)$ and $\bar{p}$ is a distribution in $\,\PP_0$  such that $\,\sum_{i=1}^{+\infty} E_i[\bar{p}_1]_i\leq E\,$  then
\begin{equation}\label{main++p}
    f(\bar{p})-f(\bar{q})\leq C\varepsilon F_{\SC}((E-E_\varepsilon(\bar{p}))/\varepsilon)+Dg(\varepsilon)\leq C\varepsilon F_{\SC}(E/\varepsilon)+Dg(\varepsilon)
\end{equation}
for any  distribution $\bar{q}$ in $\,\PP_0$ such that $\mathrm{TV}(\bar{p},\bar{q})\leq \varepsilon$, where
$$
E_\varepsilon(\bar{p})=\sum_{i=1}^{+\infty} E_i[\bar{p}^{\varepsilon}_1]_i,\quad [\bar{p}^{\varepsilon}_1]_i=\sum_{i_2,...,i_n}[p_{i,i_2,...,i_n}-\varepsilon]_+\quad ([x]_+=\max\{x,0\}),
$$
and the left hand side of (\ref{main++p}) may be equal to $-\infty$.}
\end{proposition}\smallskip

\noindent\textbf{Note B:} The equivalence of (\ref{Z-cond}) and (\ref{Z-cond+}) implies that the r.h.s. of (\ref{main+p})
and (\ref{main++p}) tend to zero as $\,\varepsilon\to0$. \smallskip


\noindent\textbf{Note C:} The  quantity $E_\varepsilon(\bar{p})$
monotonically tends to $\,E(\bar{p})\doteq\sum_{i=1}^{+\infty} E_i[\bar{p}_1]_i\,$  as $\,\varepsilon\to 0$.
So, the first estimate in (\ref{main++p}) may be essentially  sharper than the second one for small $\,\varepsilon$ and $E(\bar{p})$ close to $E$.
\smallskip

If  $\overline{X}=(X_1,...,X_n)$ and $\overline{Y}=(Y_1,...,Y_n)$
are vectors of random variables with probability distributions $\bar{p}$ and $\bar{q}$ in $\mathfrak{P}_n$ such that
the random variables $X_1$ and $Y_1$ takes values in $\SC$ then the constraint $\,\sum_{i=1}^{+\infty} E_i[\bar{p}_1]_i\leq E\,$ (respectively, $\,\sum_{i=1}^{+\infty} E_i[\bar{q}_1]_i\leq E\,$) means that $\mathbb{E}(X_1)\leq E$ (respectively, $\mathbb{E}(Y_1)\leq E$).\smallskip

\begin{example}\label{main-1-p-e} It was mentioned before that the Shannon conditional entropy (equivocation) $H(X_1\vert X_2)$ defined on the set $\PP_2^1$ by the classical version of formula (\ref{ce-ext}) belongs to the class
$T_2^1(1,1)$. Since it is a nonnegative function, Proposition \ref{main-1-p}B with $\PP_0=\PP_2^1$ and Remark \ref{main-1-p-r} imply that
\begin{equation}\label{eq-1-cb}
    H(X_1\vert X_2)_{\bar{p}}-H(Y_1\vert Y_2)_{\bar{q}}\leq \varepsilon \ln \vert \bar{p}_1\vert +g(\varepsilon)
\end{equation}
for any distribution  $\bar{p}$ in $\,\mathfrak{P}_2^1$ with finite $\vert \bar{p}_1\vert $ and
\emph{arbitrary} distribution  $\bar{q}$ in $\,\mathfrak{P}_2^1$ such that $\mathrm{TV}(\bar{p},\bar{q})\leq \varepsilon$.

It follows from (\ref{eq-1-cb}) that
\begin{equation}\label{eq-1-cb+}
-(\varepsilon \ln \vert \bar{q}_1\vert +g(\varepsilon))\leq H(X_1\vert X_2)_{\bar{p}}-H(Y_1\vert Y_2)_{\bar{q}}\leq \varepsilon \ln \vert \bar{p}_1\vert +g(\varepsilon)
\end{equation}
for any  distributions $\bar{p}$ and  $\bar{q}$ in $\,\mathfrak{P}_2^1$ such that $\,\vert \bar{p}_1\vert ,\vert \bar{q}_1\vert <+\infty\,$ and $\mathrm{TV}(\bar{p},\bar{q})\leq \varepsilon$.

If $\bar{p}$ and  $\bar{q}$ are 2-variate probability  distributions such that $[\bar{p}_1]_i=[\bar{q}_1]_i=0$ for all $i>n$ and
$\mathrm{TV}(\bar{p},\bar{q})\leq \varepsilon$ then
(\ref{eq-1-cb+}) implies that
\begin{equation}\label{my-CB}
\vert H(X_1\vert X_2)_{\bar{p}}-H(Y_1\vert Y_2)_{\bar{q}}\vert \leq \varepsilon \ln n+g(\varepsilon)
\end{equation}
(the classical version of Winter's continuity bound for the quantum conditional entropy \cite{W-CB}),
while the optimal continuity bound for  the Shannon conditional entropy obtained by Alhejji and  Smith in \cite{A&G} claims that
\begin{equation}\label{opt-CB}
\vert H(X_1\vert X_2)_{\bar{p}}-H(Y_1\vert Y_2)_{\bar{q}}\vert \leq \varepsilon \ln (n-1)+h_2(\varepsilon)
\end{equation}
provided that $\varepsilon\leq1-1/n$.

It is clear that (\ref{opt-CB}) is sharper than (\ref{my-CB}) but the difference is not too large
for $n\gg1$ and small $\varepsilon$. At the same time, continuity bound (\ref{eq-1-cb+}) may give more accurate
estimates for the quantity $\,H(X_1\vert X_2)_{\bar{p}}-H(Y_1\vert Y_2)_{\bar{q}}\,$ than continuity bound (\ref{opt-CB}).

Assume that the random variables $X_1$ and $Y_1$ takes the values $0,1,2,..$.
Proposition \ref{main-p}B  with $\PP_0=\PP_2^1$ and Remark \ref{main-1-p-r} imply that
\begin{equation}\label{eq-2-cb}
    H(X_1\vert X_2)_{\bar{p}}-H(Y_1\vert Y_2)_{\bar{q}}\leq \varepsilon g(E/\varepsilon)+g(\varepsilon)
\end{equation}
for any  distributions $\bar{p}$ and $\bar{q}$ in $\,\mathfrak{P}_2^1$ such that  $\mathbb{E}(X_1)\doteq\sum_{i=1}^{+\infty} (i-1)\shs[\bar{p}_1]_i\leq E$ and $\mathrm{TV}(\bar{p},\bar{q})\leq \varepsilon$.

To show the accuracy of the semi-continuity bound (\ref{eq-2-cb}) take any $E>0$ and $\varepsilon\in[0,E/(E+1)]\,$ and consider the probability distributions $\bar{p}=\{r_it_j\}_{i,j\geq1}$ and $\bar{q}=\{s_it_j\}_{i,j\geq1}$,
where $\{t_j\}_{j\geq1}$ is any probability distribution, while $\{r_i\}_{i\geq1}$ and $\{s_i\}_{i\geq1}$ are the probability distributions such that
$$
H(\{r_i\})=Eh_2(\varepsilon/E)+h_2(\varepsilon),\quad \sum_{i=1}^{+\infty}(i-1)r_i=E,\quad H(\{s_i\})=0,\quad  \sum_{i=1}^{+\infty}(i-1)s_i=0
$$
and $\,\mathrm{TV}(\{r_i\},\{s_i\})=\varepsilon\,$ constructed after Theorem 2 in \cite{BDJ}. Then $\mathbb{E}(X_1)=E$, $\mathbb{E}(Y_1)=0$, $\,\mathrm{TV}(\bar{p},\bar{q})=\varepsilon\,$ and
\begin{equation*}
    H(X_1\vert X_2)_{\bar{p}}-H(Y_1\vert Y_2)_{\bar{q}}=H(\{r_i\})-H(\{s_i\})=Eh_2(\varepsilon/E)+h_2(\varepsilon).
\end{equation*}
We see that in this case the l.h.s. of (\ref{eq-2-cb}) is very close to the r.h.s. of (\ref{eq-2-cb}) equal to
$$
(E+\varepsilon)h_2\!\left(\frac{\varepsilon}{E+\varepsilon}\right)+(1+\varepsilon)h_2\!\left(\frac{\varepsilon}{1+\varepsilon}\right)
$$
provided that $\varepsilon$ is small and $E\gg\varepsilon$.

The semi-continuity bound  (\ref{eq-2-cb}) implies that
\begin{equation*}
-(\varepsilon g(E_q/\varepsilon)+g(\varepsilon))\leq H(X_1\vert X_2)_{\bar{p}}-H(Y_1\vert Y_2)_{\bar{q}}\leq \varepsilon g(E_p/\varepsilon)+g(\varepsilon)
\end{equation*}
for any  distributions $\bar{p}$ and  $\bar{q}$ in $\,\mathfrak{P}_2^1$ such that $\mathbb{E}(X_1)\doteq\sum_{i=1}^{+\infty}(i-1)\shs[\bar{p}_1]_i\leq E_p$, $\mathbb{E}(Y_1)\doteq\sum_{i=1}^{+\infty} (i-1)\shs[\bar{q}_1]_i\leq E_q\,$ and $\,\mathrm{TV}(\bar{p},\bar{q})\leq \varepsilon$.\smallskip

The main advantage of the proposed technique consists in the possibility to obtain
semi-continuity bounds (\ref{eq-1-cb}) and (\ref{eq-2-cb}) \emph{with no constraint on the distribution} $\bar{q}$.
\end{example}\smallskip

Both statements of Proposition \ref{main-p} can be generalized to functions from the classes $T_n^m(C,D\vert \PP_0)$, $m>1$, by using Corollary \ref{main-c} in Section 3.2  via the mentioned above identification of $n$-variate probability distributions with quantum states in $\S(\H_{A_1...A_n})$ diagonizable in the basis
(\ref{c-basis}).

\subsection{Characteristics of classical states of quantum oscillators}

Assume that $A_1...A_n$ is a $n$-mode quantum oscillator and $\S_\mathrm{cl}(\H_{A_1...A_n})$ is the set of classical states -- the convex closure
of the family $\{\vert \bar{z}\rangle\langle \bar{z}\vert  \}_{\bar{z}\in\mathbb{C}^n}$ of coherent states \cite{H-SCI,IQO,Amosov}. Each state
$\rho$ in $\S_\mathrm{cl}(\H_{A_1...A_n})$ can be represented as
\begin{equation}\label{P-rep}
 \rho=\int_{\mathbb{C}^n}\vert \bar{z}\rangle\langle \bar{z}\vert  \mu_{\rho}(dz_1...dz_n),\quad  \bar{z}=(z_1,...,z_n),
\end{equation}
where $\mu_{\rho}$ is a Borel probability measure on $\mathbb{C}^n$.  Representation (\ref{P-rep}) is called Glauber– Sudarshan $P$-representation and generally is written as
\begin{equation*}
 \rho=\int_{\mathbb{C}^n}\vert \bar{z}\rangle\langle \bar{z}\vert  P_{\rho}(\bar{z})dz_1...dz_n,
\end{equation*}
where $P_{\rho}$ is the $P$-function of a state $\rho$, which in this case is nonnegative and can be treated as
a generalized probability density function on $\mathbb{C}^n$ (in contrast to the standard probability density,
the function $P_{\rho}$ may be singular, since the measure $\mu_{\rho}$ may be not absolutely continuous w.r.t. the Lebesgue measure on
$\mathbb{C}^n$) \cite{Gla,Sud}.


Representation (\ref{P-rep}) means that $\S_\mathrm{cl}(\H_{A_1...A_n})=\mathfrak{Q}_{X,\F,\tilde{\omega}}$ -- the set defined in (\ref{q-set}), where $X=\mathbb{C}^n$, $\F$ is the Borel $\sigma$-algebra on $\mathbb{C}^n$ and $\tilde{\omega}(\bar{z})=\vert \bar{z}\rangle\langle \bar{z}\vert$.

Let $H_{A_k}=\hat{N}_k\doteq a^\dag_ka_k$ be the number operator of the $k$-th mode, $k=\overline{1,m}$, $m\leq n$. Then the operator $H_{m}$ defined in (\ref{H-m}) is the total number operator for the subsystem $A_1...A_m$ and the function $F_{H_{m}}$ defined in (\ref{F-H-m}) has the form
\begin{equation*}
F_{H_{m}}(E)=mF_{\hat{N}_1}(E/m)=mg(E/m),
\end{equation*}
where $g$ is the function defined in (\ref{g-def}).

Denote by
$\rho(\mu)$ the classical state having representation (\ref{P-rep}) with a given measure $\mu$ in $\P(\mathbb{C}^n)$. It is easy to see that
\begin{equation}\label{N-exp}
\Tr \hat{N}_k[\rho(\mu)]_{A_k}=\int_{\mathbb{C}} \vert z\vert^2 \mu_k(dz),\quad k=1,2,..,n,
\end{equation}
where $\mu_k$ is the marginal measure of $\mu$ corresponding to the $k$-th component of $\bar{z}$, i.e. $\mu_k(A)=\mu(\mathbb{C}_1\times...\times\mathbb{C}_{k-1}\times A\times \mathbb{C}_{k+1}\times...\times\mathbb{C}_{n})$ for any $A\subseteq\mathbb{C}$.\smallskip

By applying Theorem \ref{main} in Section 3.2 we  obtain the following\smallskip

\begin{proposition}\label{main-cs} \emph{Let $\S_0$ be a convex subset of $\S(\H_{A_1...A_n})$ with the property
\begin{equation}\label{S-prop-c}
  \rho\in\S_0\cap\S_\mathrm{cl}(\H_{A_1...A_n})\quad \Rightarrow \quad\{\sigma\in\S_\mathrm{cl}(\H_{A_1...A_n})\,\vert \,\exists \varepsilon>0:\varepsilon\sigma\leq \rho\}\subseteq\S_0
\end{equation}
and $\,\P_{\S_0}(\mathbb{C}^n)\,$ the subset of $\,\P(\mathbb{C}^n)\,$ consisting of measures $\mu$ such that $\rho(\mu)\in\S_0$.}\smallskip

\emph{\noindent\emph{A)} If $f$ is a function in $\widehat{L}_n^{m}(C,D\vert \S_0)$ then
\begin{equation*}
    \vert f(\rho(\mu))-f(\rho(\nu))\vert \leq C\varepsilon mg(E/\varepsilon)+Dg(\varepsilon)
\end{equation*}
for any measures $\mu$ and $\nu$ in $\,\P_{\S_0}(\mathbb{C}^n)\,$ such that $\,\sum_{k=1}^{m}\int_{\mathbb{C}}\vert z\vert^2\mu_k(dz)\leq mE$,\\ $\sum_{k=1}^{m}\int_{\mathbb{C}} \vert z\vert^2 \nu_k(dz)\leq mE\,$ and $\,\mathrm{TV}(\mu,\nu)\leq \varepsilon$.}\smallskip

\emph{\noindent\emph{B)} If $f$ is a nonnegative function in $\widehat{L}_n^{m}(C,D\vert \S_0)$ and $\mu$ is a measure in $\,\P_{\S_0}(\mathbb{C}^n)\,$ such that $\,\sum_{k=1}^{m}\int_{C}\vert z\vert^2\mu_k(dz)\leq mE\,$  then
\begin{equation}\label{main++cs}
    f(\rho(\mu))-f(\rho(\nu))\leq C\varepsilon mg(E/\varepsilon)+Dg(\varepsilon)
\end{equation}
for any measure $\nu$ in $\,\P_{\S_0}(\mathbb{C}^n)\,$ such that $\mathrm{TV}(\mu,\nu)\leq \varepsilon$ (the left hand side of (\ref{main++cs}) may be equal to $-\infty$).}
\end{proposition}\smallskip

\begin{example}\label{main-cs-e} Consider two-mode quantum oscillator $A_1A_2$. Let $f=I(A_1\!:\!A_2)$ be the quantum mutual
information defined in (\ref{mi-d}). This is the function on the whole space $\S(\H_{A_1A_2})$  taking values in $[0,+\infty]$, which belongs to the class
$L_2^1(1,2\vert \S_\mathrm{cl})$. This follows from the inequalities (\ref{MI-LAA-1}) and (\ref{MI-LAA-2}) and  the remark after upper bound (\ref{MI-UB}), since
all the states in $\S_\mathrm{cl}$ are separable. Thus, Proposition \ref{main-cs}B with $\S_0=\S_\mathrm{cl}$ implies that
\begin{equation}\label{MI-C}
    I(A_1\!:\!A_2)_{\rho(\mu)}-I(A_1\!:\!A_2)_{\rho(\nu)}\leq \varepsilon g(E/\varepsilon)+2g(\varepsilon)
\end{equation}
for any measure $\mu$ in $\,\P(\mathbb{C}^2)\,$ such that $\,\int_{C}\vert z\vert^2\mu_1(dz)\leq E\,$ and arbitrary measure $\nu$ in $\,\P(\mathbb{C}^2)$ such that $\mathrm{TV}(\mu,\nu)\leq \varepsilon$. The condition $\,\int_{C}\vert z\vert^2\mu_1(dz)\leq E\,$ can be replaced by the symmetrical condition
$\,\int_{C}\vert z\vert^2\mu_1(dz)+\int_{C}\vert z\vert^2\mu_2(dz)\leq 2E\,$ by noting that the function $f=I(A_1\!:\!A_2)$ also belongs to the class
$L_2^2(1/2,2\vert\S_\mathrm{cl})$. This follows from the inequality
$$
I(A_1\!:\!A_2)_{\rho}\leq \textstyle\frac{1}{2}(S(\rho_{A_1})+S(\rho_{A_2}))
$$
valid for any separable state $\rho$ in $\S(\H_{A_1A_2})$, which can be proved easily by using  upper bound (\ref{MI-UB}), the remark after it and the symmetry arguments.

\smallskip

Continuity bound (\ref{MI-C}) complements semi-continuity bound (\ref{I-1-2}) for commuting states. Note that
universal semi-continuity bound for the quantum mutual information has not been constructed yet (as far as I know).

The semi-continuity bound (\ref{MI-C}) implies that
\begin{equation}\label{MI-C++}
\vert I(A_1\!:\!A_2)_{\rho(\mu)}-I(A_1\!:\!A_2)_{\rho(\nu)}\vert \leq \varepsilon g(E/\varepsilon)+2g(\varepsilon)
\end{equation}
for any measures $\mu$ and $\nu$ in $\,\P(\mathbb{C}^2)\,$ such that $\,\int_{C}\vert z\vert^2\mu_1(dz),\;\int_{C}\vert z\vert^2\nu_1(dz)\leq E\,$ and $\mathrm{TV}(\mu,\nu)\leq \varepsilon$.


The advantage of continuity bound (\ref{MI-C++}) is its simplicity  and accuracy in contrast to the universal continuity bounds for QMI under the energy constraint
(described in \cite[Section 4.2.2]{QC}). Its obvious drawback is the use of the total variation distance between representing measures as a quantity describing closeness of classical states (instead of the trace norm distance).
\end{example}

If the measures $\mu$ and $\nu$ representing classical states $\rho$ and $\sigma$ are absolutely continuous w.r.t. the Lebesgue measure on
$\mathbb{C}^n$ then $\mathrm{TV}(\mu,\nu)=\frac{1}{2}\|P_\rho-P_\sigma\|_{L_1}$ -- the $L_1$-norm distance between the $P$-functions of $\rho$ and $\sigma$.

\section{Remarks on other applications}

In this article we proposed a modification of the Alicki-Fannes-Winter technique designed for quantitative
continuity analysis of locally almost affine functions  on convex sets of states called "quasi-classical" that can be represented
as the set $\mathfrak{Q}_{X,\F,\tilde{\omega}}$ defined in (\ref{q-set}) by means of some measurable space $\{X,\F\}$ and a $\F$-measurable $\S(\H)$-valued function $\tilde{\omega}(x)$ on $X$. We have used "quasi-classical" sets of two types:
\begin{itemize}
  \item  the set of all states in $\S(\H)$ diagonizable in a particular basic in $\H$;
  \item  the set of classical states of a multi-mode quantum oscillator.
\end{itemize}

The scope of potential applications of the proposed method can be expanded using the following\smallskip
\begin{lemma}\label{mu-app}
\emph{Let $\S_0$ be a closed convex subset of $\S(\H)$. Then
\begin{equation}\label{S-rep}
\S_0=\mathfrak{Q}_{\mathrm{cl}(\mathrm{ext}\S_0),\B,\shs \id},
\end{equation}
where $\mathrm{cl}(\mathrm{ext}\S_0)$ is the closure of the
set of extreme points of $\S_0$,  $\B$ is the Borel\break $\sigma$-algebra on $\,\mathrm{cl}(\mathrm{ext}\S_0)\,$ and $\,\id$ is the identity map on $\,\mathrm{cl}(\mathrm{ext}\S_0)$.}\smallskip

\emph{The set $\,\mathrm{cl}(\mathrm{ext}\S_0)$ in (\ref{S-rep}) can be replaced by any closed subset $\,\S_*$ of $\,\S_0$
the convex closure of which  coincides with $\S_0$.}\footnote{The convex closure of a set $S$ in a Banach space is the minimal closed convex set containing $S$.}
\end{lemma}

\begin{proof} Both claims of the lemma follow from the $\mu$-compactness of the set $\S(\H)$  in terms of \cite{P&Sh}. This property
of $\S(\H)$ established in \cite[Proposition 2]{H-Sh-2} can be formulated as
\begin{equation}\label{m-comp}
\!\left\{\S\textrm{ is a compact subset of }\S(\H) \right\}\,\Leftrightarrow\,\left\{\P_{\S}\textrm{ is a compact subset of }\P(\S(\H)) \right\},\!
\end{equation}
where $\P(\S(\H))$ is the set of all Borel probability measures on $\S(\H)$ equipped with the
topology of weak convergence and $\P_\S$ is the subset of $\P(\S(\H))$ consisting of measures with
the barycenter in $\S$ \cite{Bil+,Bil}. The nontrivial implication in (\ref{m-comp}) is $"\Rightarrow"$, since $"\Leftarrow"$
follows from continuity of the barycenter map w.r.t. the weak convergence.\smallskip

The $\mu$-compactness of $\S(\H)$ allows us to
prove for this \emph{noncompact} set some general results valid for compact sets, in particular,
several results from the Choquet theory.\footnote{Another corollaries of the $\mu$-compactness of $\S(\H)$ are presented in the first part of \cite{EM}.} For example, the $\mu$-compactness of $\S(\H)$ implies, by
Proposition 5 in \cite{P&Sh}, that
\begin{equation}\label{S-rep+}
\S_0=\{\mathrm{b}(\mu)\,\vert \, \mu\in\P(\mathrm{cl}(\mathrm{ext}\S_0))\},
\end{equation}
where $\mathrm{b}(\mu)\doteq\int\rho\mu(d\rho)$ is the barycenter of $\mu$ and $\P(\mathrm{cl}(\mathrm{ext}\S_0))$ is the set of all
Borel probability measures on $\mathrm{cl}(\mathrm{ext}\S_0)$.  This proves (\ref{S-rep}).

To prove the last claim of the lemma we have to show that (\ref{S-rep+}) holds with $\mathrm{cl}(\mathrm{ext}\S_0)$ replaced by $\S_*$.
Assume that $\rho_0$ is a state in $\S_0$. Then there is a sequence $\{\rho_n\}$ from the convex hull of $\S_*$
converging to $\rho_0$. It means that $\rho_n=\mathrm{b}(\mu_n)$ for all $n$, where $\mu_n$ is a measure in $\P(\S_*)$
with finite support. Since the set $\{\rho_n\}\cup\{\rho_0\}$ is compact, the implication $"\Rightarrow"$ in (\ref{m-comp})
shows that the sequence $\{\mu_n\}$ is relatively compact in the topology of weak convergence and hence has a partial limit
$\mu_*$. The continuity of the barycenter map implies that $\rho_0=\mathrm{b}(\mu_*)$.
\end{proof}

Lemma \ref{mu-app} claims that any state $\rho$ in a closed convex subset $\S_0$ of $\S(\H)$ can be represented as
\begin{equation}\label{b-rep}
  \rho=\mathrm{b}(\mu),\quad \mu\in \P(\mathrm{cl}(\mathrm{ext}\S_0)).
\end{equation}

If $\S_0$ is the set of all states in $\S(\H)$ diagonizable in a  basic $\{\vert n\rangle\langle n\vert \}_{n=0}^{+\infty}$ in $\H$ then
$\mathrm{cl}(\mathrm{ext}\S_0)=\mathrm{ext}\S_0=\{\vert n\rangle\langle n\vert \}_{n=0}^{+\infty}\,$ and representation (\ref{b-rep})
is the spectral decomposition of $\rho$.

If $\S_0$ is the set of classical states of a multi-mode quantum oscillator then $\,\mathrm{cl}(\mathrm{ext}\S_0)=\mathrm{ext}\S_0=\{\vert \bar{z}\rangle\langle \bar{z}\vert  \}_{\bar{z}\in \mathbb{C}^n}$ and (\ref{b-rep}) is the equivalent form of the $P$-representation (\ref{P-rep}).\smallskip

As a nontrivial application of Lemma \ref{mu-app} one can consider the case when $\S_0$ is the set of separable (non-entangled) states of a
bipartite quantum system $AB$ defined as the convex closure of the set of product states of $AB$. In this case
$\mathrm{cl}(\mathrm{ext}\S_0)=\mathrm{ext}\S_0$ is the set of all pure product states and representation (\ref{b-rep}) means that
\emph{any separable state can be represented as a "continuous convex mixture" of pure product states} despite the existence of
separable states that can not be represented as a countable (discrete) convex mixture of pure product states \cite{H-Sh-W}.

By Lemma \ref{mu-app} we may apply the technique developed in this article for quantitative
continuity analysis of locally almost affine functions on any closed convex subset $\S_0$ of $\S(\H)$ using  the total variation distance between representing measures
as a quantity describing  closeness of the states in $\S_0$ (as it was made for the set of classical states of a quantum oscillator in Section 4.5).
\bigskip

I am grateful to A.S.Holevo and G.G.Amosov for valuable discussion. I am also grateful to L.Lami for the useful reference.
Special thanks to A.Winter for the  comment concerning Mirsky's inequality. I am grateful to the unknown reviewers for valuable suggestions and typos found.
This work was funded by Russian Federation represented by the Ministry of Science
and Higher Education (grant number 075-15-2020-788).

\end{document}